\documentstyle[aps,amsfonts,graphicx,eqsecnum]{revtex}

\newcommand{\sm}{{\mathsf s}}

\newcommand{\zR}{E_R - i\Gamma_R/2}

\newcommand{\HS}{\text{${\mathcal H}$}}
\newcommand{\Phx}{\text{${\mathbf\Phi}^\times$}}
\newcommand{\bk}[2]{\text{$\langle #1|#2 \rangle$}}
\newcommand{\kt}[1]{\text{$|#1\rangle$}}
\newcommand{\br}[1]{\text{$\langle #1|$}}

\newcommand{\mkt}[1]{\text{$|#1{}^{-}\rangle$}}

\begin{document}
\draft

\title{\bf Time Asymmetric Boundary Conditions and the Definition\\ of Mass
and Width for Relativistic Resonances}
\author{A.~R.~Bohm\thanks{E-mail: \texttt{bohm@physics.utexas.edu}},
R.~de la Madrid\thanks{Permanent address:
Departamento de F\'\i sica Te\'orica, Universidad de Valladolid,
47011~Valladolid, Spain}, B.A.~Tay\thanks{E-mail: \texttt{batay@physics.utexas.edu}},}
\address{Center for Particle Physics, the University of 
Texas at Austin, Austin, TX 78712-1081, USA}
\author{P.~Kielanowski\thanks{also at the Institute of
Theoretical Physics, University of Bia{\l}ystok, Poland}}
\address{Departamento de F\'\i sica, Centro de Investigaci\'on
y de Estudios Avanzados del IPN, M\'exico D.F., M\'exico}

\date{\today}
\maketitle

\begin{abstract}
The definition of mass and width of relativistic resonances and in particular
of the $Z$-boson is discussed.  For this we use the theory based on time
asymmetric boundary conditions given by Hardy class spaces ${\mathbf \Phi}_-$
and ${\mathbf \Phi} _+$ for prepared in-states and detected out-states
respectively, rather than time symmetric Hilbert space theory.  This Hardy
class boundary condition is a mathematically rigorous form of the 
singular Lippmann-Schwinger equation.  In addition to the rigorous definition
of the Lippmann-Schwinger kets $|[j,{\mathsf s}]^{\pm}\rangle$ as functionals
on the spaces ${\mathbf \Phi} _{\mp}$, one obtains Gamow kets 
$|[j,{\mathsf s}_R]^- \rangle$ with complex centre-of-mass energy value
${\mathsf s}_R=(M_R-i\Gamma _R/2)^2$.  The Gamow kets have an exponential time
evolution given by $\exp{(-iM_Rt-\Gamma _Rt/2)}$ which suggests that
$(M_R,\Gamma _R)$ is the right definition of the mass and width of a 
resonance.  This is different from the two definitions of the $Z$-boson
mass and width used in the Particle Data Table and leads to a numerical value 
of $M_R=(91.1626\pm 0.0031) \ {\rm GeV}$ from the $Z$-boson lineshape data. 
    
\end{abstract}


\section{Introduction}
\label{sec:introduction}

The Review of Particle Properties~\cite{CASO} gives two 
definitions of the mass and width of the $Z$-boson and lists two different 
values which are obtained from the fit of two different formulas for the 
lineshape to the same experimental data.  The value $M_Z$ is obtained from 
the fit to the ``relativistic Breit-Wigner with energy dependent width'' of 
the on-shell renormalisation scheme
\begin{equation}
      a_j^{om}({\mathsf s})=
     \frac{-\sqrt{\mathsf s}\sqrt{\Gamma_e({\mathsf s})\Gamma_f({\mathsf s})}}
     {{\mathsf s}-M^2_Z+i\sqrt{\mathsf s}\Gamma _Z(\mathsf s)}
     \approx 
     \frac{-M_ZB_{ef}\Gamma_Z}{{\mathsf s}-M^2_Z+i
      \frac{\mathsf s}{M_Z}\Gamma_Z}
      =\frac{R_Z}{{\mathsf s}-M^2_Z+i\frac{\mathsf s}{M_Z}\Gamma_Z} \, , \quad 
      m_0^2\leq {\mathsf s}<\infty \, .
     \label{ajom1}
\end{equation}
The value $\bar M_Z$ is obtained from the relativistic Breit-Wigner of the 
$S$-matrix pole
\begin{equation}
      a_j^{BW}({\mathsf s})=
      \frac{R_Z}{{\mathsf s}-{\mathsf s}_R}=
      \frac {R_Z}{{\mathsf s}-\bar M^2_Z+i\bar M_Z\bar\Gamma_Z}=
      \frac {R_Z}{{\mathsf s}-(M_R-i\frac{\Gamma_R}{2})^2}\, , \quad 
      m_0^2\leq {\mathsf s}<\infty \, .
      \label{ajBW2}
\end{equation}
Both lineshape formulas (\ref{ajom1}) and (\ref{ajBW2}) fit the experimental 
data  equally well~\cite{RIEMANNa,RIEMANNb,RIEMANNc}.  But they lead to 
values of the mass 
parameters $M_Z$ and $\bar M_Z$ which differ from each other by about 10 
times the experimental error.  From $a_j^{om}({\mathsf s})$ 
the fit gives the values 
\begin{equation}
      M_Z=(91.1871\pm 0.0021) \ {\rm GeV}\, , \quad 
      \Gamma_Z=(2.4945\pm 0.0024) \ {\rm GeV} \, .
      \label{MZvalue}
\end{equation}
From $a_j^{BW}({\mathsf s})$ one obtains the values 
\begin{equation}
       M_R=(91.1626\pm 0.0031) \ {\rm GeV}\, , \quad 
      \Gamma_R=(2.4934\pm 0.0024)\ {\rm GeV} \, .
       \label{MRvalue}
\end{equation}
Numerically this means 
\begin{equation}
      M_R=M_Z-0.026 \ {\rm GeV}=M_Z-10\times \Delta m_{exp}\, ,\quad 
      \Gamma_R=\Gamma_Z-1.2 \ {\rm MeV} \, ,
      \label{4a}
\end{equation}
or
\begin{equation}
       \bar M_Z=M_Z-34 \ {\rm MeV} \, . 
       \label{4b}
\end{equation}
The question thus is: What is the right definition of the $Z$-boson mass 
and width and therefore the right numerical value of the mass 
of the $Z$-boson? 

Even if one chooses the $S$-matrix definition (\ref{ajBW2}) because it is 
gauge invariant~\cite{SIRLIN,STUART}, the complex parameter 
${\mathsf s}_R$ in Eq.~(\ref{ajBW2}) can be 
expressed in terms of the real parameters mass and width in many different 
ways leading to many more arbitrary definitions of the $Z$-boson mass.  Some 
of these mentioned in the literature are
\begin{enumerate}
    \item $(\bar M_Z, \bar\Gamma_Z)$ (also called $(m_2,\Gamma_2)$
    \cite{SIRLIN})
   \begin{equation} 
        {\mathsf s}_R=\bar M_Z^2-i\bar M_Z\bar \Gamma_Z \, .
       \label{5a}
   \end{equation}
   \item $(M_R,\Gamma _R)$ \cite{WILLENBROCK} which is often used but not 
  dictated by the analytic $S$-matrix theory (see Ref.\cite{EDEN}, in particular
  p.~248)
  \begin{equation}
     {\mathsf s}_R=\left( M_R-i\frac{\Gamma_R}{2}\right)^2 \, .
     \label{5b}
  \end{equation}
     This is related to $(\bar M_Z,\bar\Gamma_Z)$ by the algebraic identity
  \begin{equation}
     \bar M_Z=M_R\left(1-\frac{1}{4}
        \left( \frac{\Gamma_R}{M_R}\right)^2\right)^\frac{1}{2}
     \quad  {\rm and}\quad 
     \bar\Gamma_Z=\Gamma_R\left(1-\frac{1}{4}\left
        (\frac{\Gamma_R}{M_R}\right)^2\right)^{-\frac{1}{2}}.
      \label{5c}
  \end{equation}
  \item  $(m_1,\Gamma _1)$ \cite{SIRLIN} which is numerically very close 
  to the conventional Standard Model values $(M_Z,\Gamma _Z)$.  It
   can be defined in terms of $\bar M _Z$ and $\bar \Gamma _Z$ by:
  \begin{equation}     
        \bar M_Z^2=m_1^2 
     \left( 1+\left(\frac{\Gamma_1}{m_1}\right)^2\right)^{-1}
      \quad {\rm and} \quad  
     \bar\Gamma_Z^2=\Gamma_1^2
     \left(1+\left(\frac{\Gamma_1}{m_1}\right)^2\right)^{-1},
     \label{5d}
   \end{equation}
  such that $m_1$ is numerically the same as $M_Z$ of (\ref{MZvalue}), if one 
  identifies the maximum of $|a_j^{om}({\mathsf s})|^2$, which is 
  $M_Z\left(1+\left(\Gamma_Z/M_Z\right)^2\right)^{-1}$, with the 
  maximum of $|a_j^{BW}({\mathsf s})|^2$, which is $\bar M_Z^2$, 
  \begin{equation}
        m_1\approx M_Z \, , \quad  \Gamma_1\approx\Gamma_Z \, .
        \label{5e}
   \end{equation}
\end{enumerate}  

Before we give an answer to the question of how to define the $Z$-boson mass 
and width, we should like to make the following remarks:  With the present 
data it is totally irrelevant which 
definition of the $Z$-boson mass one uses for a global fit of the Standard 
Model, since other Standard Model parameters such as $M_W$ or 
$1/\alpha_{EM}(M_Z)$ have errors which far exceed the differences (\ref{4a}) 
and (\ref{4b}) \cite{RIEMANNa,RIEMANNb,RIEMANNc}.  The global fit is good 
($\chi^2\approx 12$ 
for 12 degrees of freedom), confirming to the present level 
of accuracy the experimental data and also the Standard Model as a theory, 
including electroweak radiative corrections.  The exact definition 
of the $Z$-boson mass (and other resonance masses) is not a central 
issue of the Standard Model, and the prevailing opinion is that there 
are many ways to parameterize the experimental data of electroweak 
physics \cite{STUART,STUARTW} and $m_1$ may be as good a value for the 
$Z$-boson mass as $\bar M_Z$ and $M_R$.  However, though not presently needed 
for the Standard Model fits, the extraordinary accuracy of the 
$Z$-lineshape data, obtained with great expense of time and effort, 
allowed for the first time to discuss the problem of the definition 
of mass and width for an unstable relativistic particle and go beyond 
the level of precision given by the Weisskopf-Wigner approximation 
and one-loop effects.  This has opened up a new tier of inquiries, as 
witnessed by~\cite{SIRLIN,STUART,WILLENBROCK}, and led us to the 
question: what is a relativistic unstable particle?     

Our answer to this question is given, in analogy to 
the case of stable relativistic particles, in terms of 
representations of transformations of relativistic space-time 
(Poincar\'e transformations).  The states of stable 
elementary particles are vectors of an irreducible 
representation space $[m^2 , j]$ of the Poincar\'e group 
${\cal P}$~\cite{WIGNER} 
from which one can define the fields \cite{WEINBERG}.  This should not be 
restricted to interaction free, asymptotic states, but apply 
also to the exact states and to Poincar\'e transformations generated 
by the (interaction-incorporating) ``exact generators'' $P_0=H=H_0+V$, 
$P^i$, $J^{\mu \nu}$~\cite{WEINBERG}\footnote{For the conventional quantities 
we follow here fairly closely the notation of Ref.~\cite{WEINBERG}, but
omit the letters $\Psi$ and use Dirac's notation instead; thus 
$|\alpha ^{\pm}\rangle =\Psi _{\alpha}^{\pm}$ of~\cite{WEINBERG}.}.  
 
It has been claimed~\cite{WEINBERG} that the exact in-states 
$|\alpha^+\rangle$ and the out-states $|\alpha^- \rangle$
(which fulfill the Lippmann-Schwinger equation) transform under
the Poincar\'e group in the same way as the free states. 
This is not quite correct but under a precisely defined mathematical 
hypothesis, which replaces the Hilbert space axiom of standard 
quantum mechanics, one can show that the in- and out-states 
$|\alpha^+\rangle$ and $|\alpha^-\rangle$ of the Lippmann-Schwinger
equation span an irreducible representation space of 
Poincar\'e-$\it semigroup$ 
transformations into the backward and forward light cone, 
respectively \cite{RGVI-III}. The {\it semigroup}\/ representations
into the forward light cone are characterized by the angular momentum
$j$ (of the partial wave) and the centre-of-mass energy square $\mathsf s$
(which we will just call energy in the following) 
where $m^2_0 \leq {\mathsf s} < \infty$. 
Thus we denote
$|\alpha^- \rangle = |[{\mathsf s}, j], b^-\rangle$
where $b$ are some degeneracy parameters (e.g.\ $j_3$ and momentum
$\bf p$; as in Wigner's basis vectors for unitary representations
$[m^2,j]$ of the Poincar\'e group.) 
However, one can also take the 4-velocities 
${\hat {\bf p}}=\frac{{\bf p}}{\sqrt{\mathsf s}}=\gamma {\bf v}$;
$\gamma =\hat p^0 =\sqrt{1+{\hat{\bf p}}^2}$. 
The kets $|[{\mathsf s}, j],b=\hat{\bf p}\, ^-\rangle$, 
and therewith the semigroup representations $[{\mathsf s}, j]$,
will be analytically continued into the lower half plane, second 
Riemann sheet of the partial $S$-matrix $S_j (\sm)$ 
to the resonance pole position $\sm =\sm_R = (M_R-i\frac{\Gamma_R}{2})^2$.
The resulting semigroup representations $[\sm_R, j]$, obtained by
integrating around the resonance pole, are the 
resonance analogues of Wigner's stable particle representations $[m^2, j]$,
and we use $[\sm_R, j]$ as definition of the relativistic unstable
particle in the same way as $[m^2, j]$ serves as definition of 
the relativistic stable particle. 
From the transformation property
of the $|[\sm_R,j],b^-\rangle$ 
under the Poincar\'e semigroup transformations it follows
that the decay constant of the unstable particle is 
$1/\tau =\Gamma_R=-2{\rm Im}\sqrt{\sm_R}$.
Therefore if we want the lifetime = inverse width relation to hold
for relativistic unstable particles, then (\ref{ajBW2}) is the 
right function for the relativistic Breit-Wigner amplitude and
$(M_R, \Gamma_R)$ is the right parametrization in terms of ``mass''
and ``width''. In other words, if we define a state vector of a 
resonance with width $\Gamma_R$ by integrating the Lippmann-Schwinger
kets with a Breit-Wigner energy distribution around the pole position
at ${\mathsf s}={\mathsf s}_{R}$, then one can {\it derive}
that the lifetime of the exponential decay of these states is
$\tau=1/{\Gamma_{R}}$. In addition this is a semigroup (i.e., irreversible)
decay.

To discuss this result and to formulate the new axiom of quantum
theory from which this result follows, is the subject of this paper.
In section~\ref{sec:LShardycal} we conjecture the new mathematical
 hypothesis by which we
replace the Hilbert space axiom of quantum mechanics.
In section~\ref{sec:fSMpoGvec} we obtain the vector
description of a resonance in terms of the semigroup representations
$[\sm_R,j]$ from the resonance pole of the $S$-matrix.
In section~\ref{sec:tba} we consider the general case of 
$N$ interfering resonances
in the $j$th partial wave.  
We establish the connection between
each Breit-Wigner amplitude of the $S$-matrix $S_j(\sm )$
and the corresponding exponentially decaying
Gamow vector in a ``complex'' basis vector expansion.
To the well known background amplitude corresponds a non-exponential 
background vector in this complex basis vector expansion, which is
usually not considered (Weisskopf-Wigner approximation).    
 
\section{From the Lippmann-Schwinger equation to the Hardy class boundary
condition}
\label{sec:LShardycal}

The observed out-states $\psi ^-$ and the prepared in-states $\psi ^+$ are
(continuous) superposition of the $|\alpha ^{\pm}\rangle$:
\begin{equation}
      \psi ^{\mp}=\int \rho (\alpha ) d\alpha \,
       |\alpha ^{\mp}\rangle \langle ^{\mp}\alpha|
       \psi^{\mp}\rangle\, .
       \label{11}
       \stepcounter{equation}
       \eqnum{\theequation$\mp$}
\end{equation}
Here $\alpha$ stands for a whole collection of quantum numbers (eigenvalues
of a complete set of commuting observables c.s.c.o.), e.g., 4-momentum, spin,
particle species labels,\ldots
\begin{equation}
      \alpha =p^{\mu}_{(\alpha)}, j, j_3,n, \ldots , 
       \label{12}
\end{equation}
and the weight function $\rho (\alpha)$ (or measure 
$\rho (\alpha )d\alpha =d\mu (\alpha )$) is chosen such that 
\begin{equation}
       \langle ^{\mp}\alpha '|\psi ^{\mp}\rangle =
       \int \rho (\alpha )d\alpha 
       \langle ^{\mp}\alpha ' |\alpha ^{\mp}\rangle
       \langle ^{\mp}\alpha |\psi ^{\mp}\rangle \, ,
\end{equation}
i.e.,
\begin{equation}
      \langle ^{\mp}\alpha ' |\alpha ^{\mp}\rangle =
      \frac{1}{\rho (\alpha )}\delta (\alpha '-\alpha ) \, .
      \label{13}
\end{equation}
Equation (\ref{11}) is Dirac's basis vector expansion or completeness
relation (Nuclear Spectral theorem of mathematics).  The labels $\mp$ indicate
that the $|\alpha ^{\mp}\rangle$ are not only eigenvectors (generalized 
eigenvectors or eigenkets, since the $|\alpha ^{\pm}\rangle$ are 
functionals) of a complete system of commuting observables (c.s.c.o.), 
but that they also fulfill certain boundary conditions.  
For a scattering process, like the resonance formation 
$e^+e^-\to Z\to e^+e^-$, the
boundary conditions are usually formulated in terms of the Lippmann-Schwinger
equations \cite{L-Sch} written in its standard form as
\begin{equation}
      |\alpha ^{\pm}\rangle =|\alpha \rangle +
      \frac{1}{E_{(\alpha )}-H_0\pm i\epsilon}V|\alpha ^{\pm}\rangle =
      \left( 1+\frac{V}{E_{(\alpha )}-H\pm i\epsilon}\right)|\alpha \rangle
      = (\Omega ^{\pm})^{\times}|\alpha \rangle \, , 
      \label{14}
      \stepcounter{equation}
      \eqnum{\theequation$\pm$}
\end{equation}
where $E_{\alpha}=p_{(\alpha )}^0$.  Eqs.~(\ref{14}) are highly singular and 
mathematically ill defined expressions (like the Dirac kets were too, until
they were defined as functionals over the Schwartz space 
${\mathbf \Phi} \subset {\cal H}$).  We, therefore, 
want to formulate our boundary conditions also in terms of dense subspaces
${\mathbf \Phi} _{\mp}\subset {\cal H}$ of the Hilbert space ${\cal H}$ instead
of the singular (integral) equations (\ref{14}).  
Also, the Lippmann-Schwinger
equation singles out one momentum component $E=p^0$ in a non-covariant way;
we do not want to use the quantum numbers 
$p^0=E(=E_{e^+}+E_{e^-}$ for our resonance scattering process) but the 
centre of mass energy---or invariant mass squared 
${\mathsf s}=p_{\mu}p^{\mu}$.  We
therefore choose for the Dirac-Lippmann-Schwinger kets $|\alpha ^{\mp}\rangle$
the eigenkets
\begin{equation}
      |\alpha^\mp \rangle = |[j,{\mathsf s}],b^{\mp}\rangle  \, .
      \label{15}
      \stepcounter{equation}
      \eqnum{\theequation$\mp$}
\end{equation}
Here $j$ denotes the spin (and parity) and labels the partial amplitude 
(\ref{ajBW2}) (and (\ref{ajom1})).  For the $Z$-resonance $j=1$.  The
quantum number $\mathsf s$ is the invariant mass squared, and for the 
resonance formation process $e^+e^-\to Z\to e^+e^-$ its ``physical'' values 
are 
$(m_{e^+}+m_{e^-})^2 \equiv \sm_0 \leq {\mathsf s} < \infty$.  The continuous 
quantum numbers $b$ are the degeneracy quantum numbers of 
$[j,{\mathsf s}]$.  One
usually chooses for $b$ the momentum ${\bf p}$ and $j_3$, but we will later 
choose $b=\hat{\bf p}={\bf p}/\sqrt{\mathsf s}$, the space components of the
4-velocity. 

We now want to discuss the definition of the kets (\ref{15}). We
will start with a single stable particle space.
For a fixed isolated value of $\sm = m^2$ the Dirac  
kets $|[j,m^2],b\rangle$ span the irreducible 
unitary representation space of the Poincar\'e group ${\cal P}$ 
characterized by $[j,m^2]$,
\begin{equation}
      |[j,m^2],f \rangle= 
     \sum_{j_3}\int d\mu (b)\, |[j,m^2],j_3 b\rangle f(j_3, b) \, .
     \label{16}
\end{equation}
The coordinates $f(j_3,b)$ along the basis vectors 
$|[j,m^2],j_3 b\rangle$ (the wave functions) are labelled 
by the discrete $j_3$ which we will ignore and by continuous 
quantum numbers $b$ describing, e.g., the momentum resolution if $b={\bf p}$ 
and $d\mu (b)=d^3{\bf p}/(2p^0)$. If in place of the momentum one
chooses the space components of the 4-velocity 
$b=\hat{\bf p}={\bf p}/m$ as degeneracy
labels, then $f(j_3,\hat{\bf p}={\bf p}/m)$ describes the same
momentum resolution since $m$ is a fixed value.  

Equation (\ref{16}) is Dirac's basis vector expansion 
(Nuclear Spectral Theorem for a Rigged Hilbert Space 
${\mathbf \Phi} \subset {\mathcal H} \subset {\mathbf \Phi}^\times$ 
see Appendix) 
for the irreducible representation $[j, m^2]$ of 
${\cal P}$ and for a fixed value of $j$ and $m^2$ the vectors   
$|[j,m^2],f\rangle$ describe the state of a relativistic elementary 
particle with a momentum or 4-velocity distribution given by the function
$f(j_3,b)$. 
For the momentum wave functions 
$f(j_3,b), b={\bf p}$ or $b={\hat{\bf p}}$ 
one usually chooses functions of the 
Schwartz space ${\cal S}({\mathbb R}^3)$, ${\hat{\bf p}}\in{\mathbb R}^3$.
The vectors $|[j,m^2]f \rangle$, ($j, m^2$=fixed) are the elements 
of the abstract Schwartz space ${\mathbf \Phi}$ and the Dirac kets 
$|[j, m^2],b \rangle$ are elements of its dual ${\mathbf \Phi}^\times$ i.e.\ 
of the space of antilinear continuous functional on the space ${\mathbf \Phi}$
(see Appendix). The Rigged Hilbert space of the function spaces 
$$
{\cal S}({\mathbb R}^3) \subset {\cal L}^2({\mathbb R}^3) 
\subset {\cal S}^\times({\mathbb R}^3) 
$$                     
are equivalent to, or form a ``realization'' of, the triplet of abstract
spaces
$$
{\mathbf\Phi} \subset {\mathcal H} \subset {\mathbf \Phi}^\times
$$
The Dirac kets $|[j, m^2],b \rangle \in {\mathbf \Phi}^\times$ are thus
defined if one defines ${\mathbf \Phi}$ e.g.\ by its realization in terms
of the function space ${\cal S}({\mathbb R}^3)$ of (momentum)
wave functions $f(b)$.
We now consider the scattering states $|[j, \sm], b ^- \rangle$.

Since resonances have a definite value of spin parity (they appear in a 
partial wave $a_j({\mathsf s})$), we fix the value of $j$ (e.g., 
$j=1$ for $Z$), but we have to consider all ``physical'' values of 
$\mathsf s$, $m^2_0\leq \sm < \infty$, for the resonance 
scattering process (e.g., $e^+e^-\to Z\to e^+e^-$).  Therefore every 
out-state vector $\psi^-$ is a continuous linear combination of the 
Dirac-Lippmann-Schwinger kets 
$|[j,{\mathsf s}],b^-\rangle$, ($j=$ fixed, which we omit):
\begin{eqnarray}
     \psi^-_f&=&\int_{m_0^2}^{\infty}d{\mathsf s} \,
       |[j,{\mathsf s}],f^- \rangle \psi ^-({\mathsf s})  
\label{2.8a}  \\
      \psi^-&=&\int_{m_0^2}^{\infty}d{\mathsf s} 
\sum \!\!\!\!\!\!\!\!\int_b \
      |[j,{\mathsf s}],b^-\rangle \langle ^-b,[j,{\mathsf s}]|\psi^-\rangle
      \, ,   
 \label{2.8b}
\end{eqnarray}
where the set of admitted centre of mass energy 
wave functions $\psi^-(\sm)$ for a fixed $f$
(fixed momentum distribution $f(j_3, b)$) or in general the set of 
admitted four-momentum (or $\sm , {\hat{\bf p}}=b$) wave 
functions
\begin{equation}
\langle ^-b,[j,{\mathsf s}]|\psi ^- \rangle \equiv 
      \psi ^-(j,{\mathsf s},b) = \psi^-(\sm )
\label{2-9a}
\end{equation}
defines the space of vectors {$\psi^-$}.  
The $\psi ^-(j,{\mathsf s},b)$ 
have already been fixed to be Schwartz space functions 
of the variable $b$.  
We now want to consider the function $\psi^-(j,\sm,b)$ as functions
of $\sm$.
Since we are now interested in the variable $\mathsf s$, we often 
ignore the $b$ dependence and omit $\sum_b \!\!\!\!\!\!\!\!\int \ $ 
over the discrete and/or continuous $b$.
We therefore consider the function $\psi^-(\sm)$ for any fixed
$f \in {\mathbf\Phi}$ or the function $\psi^-(j,\sm,b)$ for any fixed values
$j, b$, which we then also call $\psi^-(\sm)=\psi^-(j,\sm,b)$.
In short we will write for (\ref{2.8a}) and (\ref{2.8b}) as:
\begin{equation}
   \psi^-=\int d\sm |\sm ^-\rangle \psi^-(\sm)
\label{2.8c}
\end{equation}
This is again the Dirac basis vector expansion or Nuclear Spectral 
theorem, but now only for the energy $\sm$, i.e.\
for the total mass-square operator 
$P_\mu P^\mu =(P_{e^+}+P_{e^-})_\mu (P_{e^+}+P_{e^-})^\mu$
whose eigenvalue is $\sm$. 
Like every Dirac ket, the ket $|\sm ^- \rangle$ is defined by
(1) the eigenvalue equation
\begin{equation}
   (P_\mu P^\mu)^\times |\sm ^-\rangle = \sm |\sm ^- \rangle
\label{2.9a}
\end{equation}
and by (2) the boundary condition
\begin{equation}
    |\sm ^-\rangle \in {\mathbf \Phi}_+^\times\,.
\label{2.9b}
\end{equation}
Usually one assumes that Dirac kets are Schwartz space functionals
like the momentum eigenkets $|b\rangle$. For the out (and in)
scattering states (\ref{15}$\mp$) that fulfill the Lippmann-Schwinger equations
(\ref{14}$\mp$) we shall make other assumptions. As already indicated 
by the label $\mp$ at the kets and $\pm$ at the spaces 
${\mathbf\Phi}^\times_{\pm}$
we shall assume different boundary conditions for the out 
and in plane wave states $|\alpha^\mp\rangle$.
To conjecture this boundary conditions we use the heuristic
Lippmann-Schwinger equations (\ref{14}$\mp$) for guidance. 

The energy wave function 
$\psi ^- ({\mathsf s})=\psi ^- (j, {\mathsf s}, b)$
considered as a function of ${\mathsf s}$ should also be a smooth, 
well behaved function. 
That means $\psi^-(\sm) \in {\cal S}({\mathbb R}_+)$
where ${\mathbb R}_+=\lbrace \sm : m^2_0 \leq \sm < \infty \rbrace$.
In the relativistic case ${\cal S}({\mathbb R}_+)$ is not exactly the
Schwartz space (since we also want $\sqrt{\sm} \psi^-(\sm)$ to be
a well behaved function), but it is a closed subspace thereof which is
dense in ${\cal L}^2({\mathbb R}_+)$. 
We will not be concerned with these mathematical details here and refer
to \cite{I-III,Sujeewa}. 

However, the functions $\psi^-(\sm)$ need to be better than well behaved,
because the $-i\epsilon$ in the Lippmann-Schwinger equation
(\ref{14}$-$) indicates that the function 
$\langle \psi ^-|\sm ^-\rangle =\overline{\langle ^-\sm |\psi ^-\rangle}
=\overline{\psi}^-({\sm })$ also needs to have some
meaning when the energy acquires a negative imaginary part.  
This we generalize by requiring that the $\overline{\psi}^-(\sm)$ can be
continued to analytic functions in the lower half complex plane that
vanish sufficiently fast at the infinite semicircle. Precisely we assume: 
\begin{equation}
      \overline{\psi}^-({\mathsf s}) \in {\cal S}\cap {\cal H}_-^2 \, ,
      \label{20-}
\end{equation}
where ${\cal H}_-^2$ is the space of Hardy functions~\cite{HARDY} of the
lower half complex plane, for which we choose the second sheet 
of the Riemann energy surface for
the analytically continued $j$-th partial $S$-matrix $S_j({\mathsf s})$.  Thus
the Lippmann-Schwinger equation suggests that its complex conjugate 
$\psi^-(\sm)$ fulfill: 
\begin{equation}
      \psi ^-({\mathsf s}) \in {\cal S}\cap {\cal H}_+^2 \, ,
      \label{20-b}
\end{equation}
where ${\cal H}_+^2$ is the space of Hardy functions~\cite{HARDY} in the
upper half complex plane. 
One can show that the space of well behaved Hardy functions 
forms a Rigged Hilbert Space (RHS) \cite{GADELLA}
\begin{equation}
      {\cal S} \cap {\cal H}^2_+ \subset {\cal L}^{2}({\mathbb R}_+)
         \subset ({\cal S} \cap {\cal H}^{2}_+)^\times
      \label{2-13}
\end{equation}
The abstract RHS whose mathematical realization is given by this triplet of 
function spaces (\ref{2-13}) we denote by:
\begin{equation}
       {\mathbf \Phi} _+ \subset {\cal H} \subset {\mathbf \Phi} _+^{\times} \, .
       \label{21-}
\end{equation}      
This means that the vectors $\psi ^-$ which
have the Dirac basis vector expansion (\ref{2.8c}) (or (\ref{2.8b})) 
with the wave functions (coordinates)
$\psi ^-({\mathsf s})\in {\cal S}\cap {\cal H}_+^2$ are elements 
$\psi ^- \in {\mathbf \Phi} _+\subset {\cal H}$.  
The basis vectors $|\sm^-\rangle \equiv|[j,{\mathsf s}], b^- \rangle$ are then 
continuous antilinear functionals on the space ${\mathbf \Phi} _+$, i.e.,
$|[j,{\mathsf s}], b^- \rangle \in {\mathbf \Phi} _+^{\times}$.  
There are more elements in ${\mathbf \Phi} _+^{\times}$ than the complete 
system of basis vectors of (\ref{2.8b}), (\ref{2.8c}).  
The elements that we are particularly
interested in are the Gamow kets $|[j,{\mathsf s}_R], b^- \rangle$ associated 
with the resonance pole of the $S$-matrix $S_j({\mathsf s})$ at 
${\mathsf s}={\mathsf s}_R$.  To generalize the boundary condition from the 
Lippmann-Schwinger equation (\ref{14}$-$) with infinitesimal imaginary
energy $-i\epsilon$ to the Hardy class spaces 
${\cal S}\cap{\cal H}^2_-$, (\ref{20-}) of the whole lower half complex plane 
may appear far fetched, 
but with the resonance pole in mind
one does not seem to have another choice.  

The Dirac-Lippmann-Schwinger kets $|[j,{\mathsf s}],b^-\rangle$ are 
generalized eigenvectors of the total mass operator 
$P_\mu P^\mu =(P_{e^+}+P_{e^-})_\mu (P_{e^+}+P_{e^-})^\mu$ 
with real eigenvalue $\mathsf s$, 
\begin{equation}
      \langle P_\mu P^\mu \psi^-|[j,{\mathsf s}],b^- \rangle ={\mathsf s}\,
      \langle \psi^-|[j,{\mathsf s}],b^-\rangle \, , \quad 
      \text{for all } \psi^- \in {\mathbf \Phi} _+ \, .
      \label{7a}
\end{equation}
This is also written as (\ref{2.9a}),
where $(P_\mu P^\mu)^{\times}$ is the extension of the adjoint operator 
$(P_\mu P^\mu)^{\dagger}=(P_\mu P^\mu)$ from the space 
${\cal H}$ to the space ${\mathbf \Phi} _+^{\times}$.  
In Dirac's 
notation the $^{\times}$ is omitted which we shall also do unless it is 
needed for clarification.

The Dirac-Lippmann-Schwinger kets at 
rest $|[j,{\mathsf s}],b_{\text{rest}}^-\rangle$ are also eigenkets 
of the full Hamiltonian $P^0\equiv H=H_0+V$:
\begin{equation}
      P^0|[j,{\mathsf s}],b_{\text{rest}}^-\rangle =
     \sqrt{\mathsf s} \, |[j,{\mathsf s}],b_{\text{rest}}^-\rangle
      \label{7c}
\end{equation}
and these kets transform under Lorentz transformations 
${\cal U}(\Lambda )$ in the well-known way 
(like the $|[j,m^2],b_{\text{rest}}\rangle$).

In the same way we wrote (\ref{11}$-$) in detail as (\ref{2.8b}) or 
as (\ref{2.8c}), we will write
(\ref{11}$+$) as continuous linear superpositions of the 
$|[j,{\mathsf s}],b\,^+\rangle$.  
Instead of choosing wave functions 
(\ref{20-b}) we choose now wave functions $\phi ^+({\mathsf s})$ which fulfill
\begin{equation}
      \phi ^+({\mathsf s})\in {\cal S}\cap {\cal H}_-^2 \, ,
      \label{20b}
\end{equation}
where ${\cal H}_-^2$ is the space of Hardy class functions on the lower half 
plane of the second sheet. 
Like in the case of (\ref{2-13}) these spaces of Hardy class functions
form another RHS \cite{GADELLA}
\begin{equation}
      {\cal S} \cap {\cal H}^2_- \subset {\cal L}^2({\mathbb R}_+) 
          \subset ({\cal S} \cap {\cal H}^2_-)^\times
\label{2-18'}
\end{equation}
where ${\cal L}^2({\mathbb R}_+)$ is the {\it same}\/ Hilbert space of Lebesgue
integrable functions as in (\ref{2-13}). The abstract RHS equivalent to 
the triplet of function spaces (\ref{2-18'}) we call
\begin{equation}
      {\mathbf\Phi}_- \subset {\cal H} \subset {\mathbf\Phi}^\times_-
\label{2-18}
\end{equation}
and ${\cal H}$ in (\ref{2-18}) and in (\ref{21-}) are the {\it same}\/ Hilbert spaces.
We denote the vector whose wave function 
is $\phi^+(\sm)$ by $\phi ^+$.  The Dirac basis vector expansion
(\ref{11}$+$) takes then the mathematically precise form
\begin{equation}
     {\mathbf \Phi} _- \ni \phi^+ = \int_{m_0^2}^{\infty}d{\mathsf s} \,
       |[j,{\mathsf s}],\phi ^+ \rangle \phi ^+({\mathsf s})
     =\int_{m_0^2}^{\infty}d{\mathsf s} \sum \!\!\!\!\!\!\!\!\int_b \
      |[j,{\mathsf s}],b^+\rangle \langle ^+b,[j,{\mathsf s}]|\phi^+\rangle
     \label{17}
\end{equation}
This is the Nuclear Spectral Theorem in the RHS (\ref{2-18}).

The heuristic Lippmann-Schwinger equations (\ref{14}$\pm$) have
thus led us to a new pair of RHS (\ref{21-}) and (\ref{2-18})
and a new pair of boundary conditions
\begin{equation}
      |[j, \sm],b\,^- \rangle \in {\mathbf\Phi}^\times_+ 
          \qquad {\rm and} \qquad 
      |[j, \sm],b\,^+ \rangle \in {\mathbf\Phi}^\times_-
\label{2-20'}
\end{equation}
for the solutions of the generalized eigenvalue equations
for the self-adjoint operator $P_\mu P^\mu$. The condition
(\ref{2-20'}) together with the required Hardy class property of the
spaces makes the heuristic Lippmann-Schwinger equations (\ref{14}$\mp$)
mathematically precise. Whether the Hardy class condition is the 
minimal requirement to accomplish this, we do not want to discuss here. 
In order to emphasise that the space of the in-states $\{ \phi ^+ \}$ 
is different from the space of the out-states $\{ \psi ^- \}$, we
changed from the notation of (\ref{11}) and denoted the in-state vectors
$\phi ^+$ by a different letter $\phi$ than the out-state vectors 
$\psi ^-$.  We want to distinguish the in-states 
$\phi ^+\in {\mathbf \Phi} _-$ and their basis vectors 
$|[j,{\mathsf s}],b^+\rangle \in {\mathbf \Phi} _-^{\times}$ 
from the out-states
$\psi ^-\in {\mathbf \Phi} _+$ and their basis vectors 
$|[j,{\mathsf s}],b^-\rangle \in {\mathbf \Phi} _+^{\times}$.
The kets $|[j,\sm],b\, ^+\rangle$ also fulfill the
eigenvalue equations (\ref{7a}),(\ref{7c}) but the two kets
fulfill different boundary
conditions expressed by the choice of the spaces 
(\ref{21-}) and (\ref{2-18}).  However note that 
${\mathbf \Phi} _-\cap {\mathbf \Phi} _+$ is not empty and it may be even 
dense in the same Hilbert space ${\cal H}$~\cite{FIND}\footnote{The
inclusion $\subset$ in (\ref{2-18}) and (\ref{21-}) must not
be understood like the inclusion of the two-dimensional space
in a 3-dimensional space but rather like the inclusion of the
rational numbers ${\mathbb Q}$ in the real numbers ${\mathbb R}$;
${\mathbb Q} \subset {\mathbb R}\,$.}.

Mathematically the space ${\mathbf \Phi} _+$ (or ${\mathbf \Phi} _-$) is defined
by the choice of function spaces 
$\left. {\cal S}\cap {\cal H}_+^2 \right| _{[m_0^{2},\infty)}$ 
(or $\left. {\cal S}\cap {\cal H}_-^2 \right| _{[m_0^{2},\infty)}$) for the 
wave functions $\psi ^-({\mathsf s})$ (or $\phi ^+({\mathsf s})$). The 
abstract vector space ${\mathbf \Phi} _{\pm}$ is mathematically ``realized''
by the function space 
$\left. {\cal S}\cap {\cal H}_{\pm}^2 \right| _{[m_0^2,\infty)}$ in
the same way as the Hilbert space ${\cal H}$ is realized by the space of 
Lebesgue square integrable functions $L^2({\mathbb R}_+,d{\mathsf s})$ and 
the abstract Schwartz space ${\mathbf \Phi}$ is realized by the space of smooth,
infinitely differentiable, rapidly decreasing functions 
${\cal S}({\mathbb R})$.  

Physically $\{ \phi^+ \}$ is the set of prepared in-states defined by the 
preparation apparatus, and $\{ \psi^- \}$ is the set of registered 
out-observables (which are usually also called out-states but) which are 
defined by the observation device (detector).  

Standard Hilbert space quantum mechanics, often including  
scattering theory, assumes 
\begin{equation}
     \{ \phi^+ \}=\{ \psi^- \}={\cal H} 
\label{2-20}
\end{equation} 
or $\{ \phi^+ \}=\{ \psi^- \}={\cal D}$,
a dense subspace of ${\cal H}$. 
If we chose, for instance, 
$\{ \phi ^+ ({\mathsf s})\}=\{ \psi ^- ({\mathsf s})\}=
{\cal S}({\mathbb R})$ (or ${\cal H}$), then
$|[j,{\mathsf s}],b_{\text{rest}}^-\rangle$ and 
$|[j,{\mathsf s}],b_{\text{rest}}^+\rangle$ 
could not fulfill the Lippmann-Schwinger 
equations (\ref{14}).
Thus there is some discrepancy between the Hilbert space axiom and 
the Lippmann-Schwinger equation.  

The axiom (\ref{2-20}) amounts to solving the time-symmetric dynamical
equations (such as the Schrodinger equation) with time-symmetric
boundary conditions. This is the only possibility within orthodox Hilbert
space quantum theory, because the Hilbert space allows only for
reversible time evolution given by the unitary group
$ U^\dagger (t)={\rm exp}(-iHt), -\infty < t < +\infty$. 
For some idealised physical states, for example the stationary states
of the harmonic oscillator, this is an acceptable boundary condition.
For scattering states this is not acceptable because
the in-state $\phi^+$ must be prepared at times
$t'$ $\it before$ a time $t_0$ $(t'<t_0)$, and the out-observable
$\psi^-$ can be registered only at times $t''$ $\it after$ $t_0$ ($t''>t_0$). 
Therefore the space of prepared in-states
should be distinguished from the space of detected out-observables.
With the new mathematical apparatus of RHS's of Hardy class one
can now distinguish between states and observables 
within the mathematical theory if one replaces the axiom (\ref{2-20}) 
by the new hypothesis:
\begin{eqnarray}
&& {\rm The\,\,space\,\,of\,\,prepared\,\,states\,\,
      |\phi^+\rangle\langle\phi^+|\,\,
        defined\,\,by\,\,the\,\,preparation\,\,apparatus\,\,is}   \nonumber \\
&&    \lbrace \phi^+ \rbrace = {\mathbf \Phi}_-\subset{\cal H}. 
                 \nonumber \\
&& {\rm The\,\,space\,\, of\,\, registered\,\, observabes}\,\,
        |\psi^-\rangle\langle\psi^-|\,\, {\rm defined\,\, by\,\,
         the\,\, registration\,\, apparatus\,\,is}   \nonumber \\
&&        \lbrace \psi^- \rbrace ={\mathbf \Phi}_+\subset{\cal H}. 
\label{2-21}
\end{eqnarray}
This new hypothesis replaces the Hilbert space hypothesis (\ref{2-20}).
The new hypothesis not only allows the continuation of the 
Lippmann-Schwinger kets $|[\sm,j]b^-\rangle\,\,$($|[\sm, j]b^+\rangle$)
into the lower (upper) half complex energy plane for which
we choose the second sheet of the $S$-matrix, but it also allows
the Gamow kets to be obtained from the resonance poles of the $S$-matrix.
This was the original motivation for the introduction of the 
Hardy functions \cite{JPM81}.

The distinction between prepared 
in-states ${\mathbf \Phi}_-$ and registered observables ${\mathbf \Phi}_+$, 
by the use of different dense (complete in a different topology than the
${\cal H}$ topology) subspaces of the $\it same$ ${\cal H}$ is the only 
modification 
in the foundation of the theory; the dynamical equations
and the algebra of observables, like the commutation relations 
of the Poincar\'e transformations,  
remain the same.    
All other novel conclusions, like the exact
exponential decay law, the precise definition of width and mass of an 
$S$-matrix pole resonance and the quantum mechanical time asymmetry 
are mathematical consequences of this new hypothesis (\ref{2-21}).

\section{From the $S$-matrix pole to the Gamow vector}
\label{sec:fSMpoGvec}

We define a resonance by a second sheet $S$-matrix pole at the complex value 
${\mathsf s}={\mathsf s}_R$.  It is thus characterized by two real 
parameters, e.g., by ${\rm Re}(\mathsf{s}_R)=\bar M_Z^2$ and 
${\rm Im}(\mathsf{s}_R)\equiv -\bar M_Z\bar\Gamma_Z$ 
as in (\ref{5a}) or by ${\rm Re}(\sqrt{{\mathsf s}_R})\equiv M_R$ and 
${\rm Im}\sqrt{{\mathsf s}_R}\equiv - \Gamma_R /2$ as in (\ref{5b}) 
or in still other ways.  The most suitable choice of this parameterisation 
of the complex ${\mathsf s}_R$ by two real numbers $(m,\Gamma)$ depends upon 
the physical interpretation of these numbers.  Since 
Weisskopf-Wigner~\cite{WW} many physicists believe that resonances and 
(exponentially) decaying states 
are the same and, especially for non-relativistic quantum mechanics, a common 
assumption is that the width $\Gamma$ of the resonance is related to the
lifetime of the decay $\tau$ by
\begin{equation}
       \frac{\hbar}{\Gamma}=\tau.
       \label{22}
\end{equation}
The width $\Gamma$ is measured as the Breit-Wigner 
line width in the cross section 
\begin{equation}
     |a^{BW}(E)|^2\sim \left|\frac{1}{E-(E_R-i\frac {\Gamma}{2})}\right|^2=
     \frac{1}{(E-E_R)^2+\frac {{\Gamma}^2}{4}} \, ,
     \label{23}
\end{equation}
and the lifetime $\tau$ is measured by the exponential law for the counting 
rate $\dot{N}_{\eta}(t)$ of the decay products $\eta$ in the decay of the
resonance $R\to \eta$:
\begin{equation}
      {\dot N}_\eta(t)\equiv \frac {\Delta N_\eta (t)}{\Delta t}\sim 
      e^{-\frac{t}{\tau}} \, .
      \label{24}
\end{equation}

Though there has not been an exact mathematical proof of
(\ref{22}) -- since the Weisskopf-Wigner method is an approximation 
(cf.~\cite{KHALFIN}) -- and though there has not been  
an accurate experimental verification of (\ref{22})\footnote{For a given 
relativistic quasistable particle one either measures 
$\Gamma$ by (\ref{23}) (for $\Gamma_R/M_R \sim 10^{-1}$) or one measures 
$\tau$ by (\ref{24}) (for $\Gamma/M \sim 10^{-10}$) and one does not
come close in accuracy to an experimental test of (\ref{22}).}, 
this relation between width and
lifetime appears to be favoured by almost everyone in relativistic 
and non-relativistic physics. 
The state vector of a quasistable state which we constructed  
in the $\it non$-relativistic 
theory~\cite{NONRELAa,NONRELAb,NONRELAc} fulfills (\ref{22}).  
We have called these quasistable state vectors, 
Gamow vectors $\psi ^G(t)$, they cannot be elements of 
the Hilbert space ${\cal H}$, but are kets on Hardy spaces.
A vector description is needed if one wants to consider the resonance of 
a formation process, like $e^+e^-\to Z \to e^+e^-$, also as a decaying
particle, like the $K_{S,L}^0 \to \pi ^+\pi ^-$ (where in the 
past one attributed
to it an eigenvector of a  complex mass or energy matrix \cite{LEE}).  

According to the fundamental assumption of quantum
mechanics, the probability for the decay $R\to \eta$ is given by the 
Born probability  
\begin{equation}
      P_\eta(t)={\rm Tr}(\Lambda_\eta |\psi^G(t)\rangle \langle \psi^G(t)|)\,,
      \label{25}
\end{equation}
where $\Lambda _{\eta}$ is the projection operator on the subspace of the 
decay products $\eta$.  The probability rate, or decay rate into $\eta$, 
should then be calculated from (\ref{25})
\begin{equation}
      \dot P_{\eta}(t)={\frac{d}{dt}}P_\eta(t) \, .
      \label{26a}
\end{equation}
Since the experimental counting rate $\dot N_{\eta}(t)$ measures the 
theoretical $\dot P_{\eta}(t)$, the latter must fulfill
\begin{equation}
      \dot P_{\eta}(t)=\Gamma _{\eta}e^{-t/\tau} \, ,
       \label{26b}
\end{equation}
in order to ensure agreement with the experimental formula (\ref{24})
with $\tau$ calculated from the property of the decaying state vector 
$\psi ^G$.

We shall construct the $\it relativistic$ Gamow vector in complete analogy
to the non-relativistic case, starting from the resonance pole of the 
$\it relativistic$ $S$-matrix at ${\mathsf s}={\mathsf s}_R$.  We shall 
see that, if ``width'' $\Gamma$ and lifetime $\tau$ are to fulfill 
(\ref{22}), then $\Gamma$ must be chosen as the parameter 
\begin{equation}
      \Gamma\equiv\Gamma_R=-2{\rm Im}(\sqrt{{\mathsf s}_R})
\end{equation}
in the relativistic Breit-Wigner formula (\ref{ajBW2}).
The real ``mass'' of the quasistable relativistic state 
will then be given by the parameter
\begin{equation}
      M_R={\rm Re}(\sqrt{{\mathsf s}_R}).
\end{equation}
Constructing a vector that fulfills (\ref{22}) starting from
another Breit-Wigner amplitude, e.g.\ (\ref{ajom1}), appears not 
possible, since (\ref{ajBW2}) plays a very special role in the 
mathematics (it is the Cauchy kernel) which is needed for the construction.

We start with the $S$-matrix element between a prepared in-state 
$\phi ^+$ and a detected out-observable 
$\psi ^-$. We assume the asymmetric boundary 
conditions $\phi ^+ \in {\mathbf \Phi} _-$ and $\psi ^- \in {\mathbf \Phi} _+$
of Section~\ref{sec:LShardycal}.
\begin{eqnarray}
      (\psi ^{out}, \phi ^{out})&=&(\psi^{out},S\phi^{in})=
      (\Omega ^- \psi ^{out},\Omega ^+ \phi ^{in})=(\psi^-,\phi^+) 
      \nonumber \\
      &=&\sum_{j,j_3,n} \int \frac{d^3\hat{\bf p}}{2\hat{E}}\, d{\mathsf s}
      \sum_{j',j_3',n '}\frac{d^3\hat{\bf p}'}{2\hat{E}'}\, d{\mathsf s}' \,
      \langle \psi ^-|[{\mathsf s},j],n, j_3, \hat{\bf p}^-\rangle \nonumber \\
      &&\times 
      \langle \hat{\bf p},j_3,[{\mathsf s},j],n |S|
          [j',{\mathsf s}'],j_3', \hat{\bf p}', n' \rangle
      \langle ^+j_3',\hat{\bf p}',[j',{\mathsf s}'],n '|\phi ^+ \rangle \, .
      \label{28}
\end{eqnarray}
In this $S$-matrix element, $\phi ^{in}$ describes the asymptotically 
free in-state that is prepared, e.g.~by the accelerator, outside the 
interaction region.  This $\phi ^{in}$ becomes the $\phi ^+$ in the 
interaction region (of the two beams in $e^+e^-\to Z\to \bar{f}f$),
the energy distribution in the beams is described by the 
wave functions $\phi ^{in}({\mathsf s})=\phi ^+({\mathsf s})$.  
The out-state 
vector $\psi ^{out}$ describes the detected out-particles (e.g., a 
particular $\bar{f}f$) when they are asymptotically free.  It comes from the
$\psi ^-$ in the interaction region and its wave function 
$\psi ^{out}({\mathsf s})=\psi ^- ({\mathsf s})$ describes the energy 
resolution of the detectors.  $\psi ^-$ is defined by the registration 
apparatus 
(detector)---for which reason $|\psi ^- \rangle \langle \psi ^-|$ should be
called observable rather than out-state.  The kets 
$|[j,{\mathsf s}],b^{\mp}\rangle$
are the eigenvectors of the exact energy operator $P_{\mu}P^{\mu}$.  The kets
$|[j,{\mathsf s}],b\rangle$ are the corresponding eigenvectors of the 
asymptotically free energy operator and 
\begin{equation}
      \psi ^{out}({\mathsf s})\equiv 
      \langle b,[j,{\mathsf s}]|\psi ^{out} \rangle
      =\langle ^-b,[j,{\mathsf s}]|\psi ^- \rangle \equiv 
       \psi ^-({\mathsf s}) \, .
\end{equation}
In addition to the property that 
$|\psi ^{out}({\mathsf s})|^2=|\psi ^-({\mathsf s})|^2$
and $|\phi ^{in}({\mathsf s})|^2=|\phi ^+({\mathsf s})|^2$ 
be a smooth function of
$\mathsf s$ (since they describe apparatus resolutions), we also require 
according to our new hypothesis of Section~\ref{sec:LShardycal} that these
functions have certain analyticity properties (Hardy class).  This is related
to causality based on the fact that the in-state $\phi ^+$ must be 
prepared first before the out-observable $\psi ^-$ can be detected in 
it~\cite{BOHM99}.  
The $S$-matrix element $|(\psi ^-,\phi ^+)|^2$ describes 
the probability to detect
the observable $\psi ^-$ in the state $\phi ^+$ (Born probability).  
This is also expressed by the asymptotically-free quantities 
$|(\psi ^{out},\phi ^{out})|^2$, where $\phi ^{out}=S\phi ^{in}$ is a state,
(not an observable like $\psi ^{out}$), which is defined by the preparation
apparatus as a $\phi ^{in}$ and the dynamics described by the $S$-operator
(or by the Hamiltonian $H$ if $S$ is calculated in terms of $H=H_0+V$).

In expressing the matrix element $(\psi ^-,\phi ^+)$ by the r.h.s.~of 
(\ref{28}), we have used for the $\phi ^+$ and $\psi ^-$ the basis vector
expansions (\ref{2.8b}) and (\ref{17}) and chosen for the quantum numbers
$b$ the space components of the 4-velocity, 
$b=\hat{\bf p}\equiv {\bf p}/\sqrt{\mathsf s}$, $j_3$ 
and $n$, where $n$ are any additional (e.g., channel) quantum numbers.  For
the Lorentz invariant integration we choose 
$d\mu (b)=d^3\hat{\bf p}/(2\hat{p}^0)$.  And we have written the $S$-matrix as 
\begin{eqnarray}
     \langle ^-\hat{\bf p},j_3,[{\mathsf s},j],n |
      \hat{\bf p}',j_3',[{\mathsf s}',j'],n'{^+} \rangle &=&
      (\Omega ^-|\hat{\bf p},j_3,[{\mathsf s},j],n \rangle, 
       \Omega ^+|\hat{\bf p}',j_3',[{\mathsf s}',j'],n' \rangle ) \nonumber \\
      &=& \langle \hat{\bf p},j_3,[{\mathsf s},j],n |\Omega {^-}{^{\dagger}} 
       \Omega ^+|\hat{\bf p}',j_3',[{\mathsf s}',j'],n' \rangle  \nonumber \\
      &=& \langle \hat{\bf p},j_3,[{\mathsf s},j],n |S|
      \hat{\bf p}',j_3',[{\mathsf s}',j'],n' \rangle \, .
      \label{29}
\end{eqnarray}
From the invariance of the $S$-operator with respect to Poincar\'e 
transformations one can show that the $S$-matrix element (\ref{29})
can be written as
\begin{equation}
      \langle \hat{\bf p},j_3,[{\mathsf s},j],n |S|
      \hat{\bf p}',j_3',[{\mathsf s}',j'],n' \rangle =
      2\hat{E}(\hat{p})\delta ^3 (\hat{p}-\hat{p}')
      \delta({\mathsf s}-{\mathsf s}')
      \delta _{j_3j_3'}\delta_{jj'} \langle n \| S_j({\mathsf s})\| n '\rangle
      \, ,
       \label{30}
\end{equation}
where $\langle n \| S_j({\mathsf s})\| n '\rangle$ is the reduced $S$-matrix
element which depends upon $j$ (which labels the partial wave; it is the
total orbital angular momentum for the case without spins, e.g., the 
$\pi ^+ \pi ^-$ system) and the particle species and channel quantum numbers
$n$, $n'$.  For a fixed initial state $n'$ it is written as
\begin{equation}
      \langle n \| S_j({\mathsf s})\| n'\rangle=S_j({\mathsf s})=
      \left\{ \begin{array}{lll}
     &2ia_j({\mathsf s})+1 \quad &\mbox{for elastic scattering $n=n '$} \\
     &2ia_j^{(n)}({\mathsf s}) \quad &\mbox{for reaction from $n'$
      into the channel $n$} \, ,
      \end{array} \right.  
     \label{31}
\end{equation}
where $a_j({\mathsf s})$ and $a_j^{(n)}({\mathsf s})$ are the partial wave
amplitudes used in (\ref{ajom1}) and (\ref{ajBW2}) (for 
$e^+e^-\to Z\to e^+e^-$ and $e^+e^- \to Z\to \mu \bar{\mu}$, 
etc.,).  We insert (\ref{30}) and (\ref{31}) into (\ref{28}) and 
obtain for the $S$-matrix element (omitting the additional quantum numbers
$n$):
\begin{equation}
      (\psi ^-,\phi ^+)=\sum_j\int_{m_0^2}^{\infty}
      d{\mathsf s} \, \sum_{j_3}\int \frac{d^3\hat{\bf p}}{2\hat{E}} 
      \langle \psi ^-|[j,{\mathsf s}],j_3, \hat{\bf p}^-\rangle
      S_j({\mathsf s}) 
      \langle ^+j_3,\hat{\bf p},[j,{\mathsf s}]|\phi ^+\rangle \, .
      \label{32}
\end{equation}
After we have made use of the Poincar\'e invariance of the $S$-matrix using
the 3-velocities basis vectors 
$|[j,{\mathsf s}],b^{\mp}\rangle =
|[j,{\mathsf s}],j_3,\hat{\bf p}^{\mp}\rangle$ we ignore
again the degeneracy quantum numbers $b$ and we consider only the $j$-th 
partial $S$-matrix element (where $j$ is the spin-parity of the resonance)
\begin{equation}
       (\psi ^-,\phi ^+)_j=\int_{m_0^2}^{\infty}d{\mathsf s} \, 
       \langle \psi ^-|{\mathsf s}^- \rangle S_j({\mathsf s})
       \langle ^+{\mathsf s}|\phi ^+ \rangle \, ,
       \label{33}
\end{equation}
where the wave functions are those of (\ref{2.8b}) and (\ref{17}),
\begin{eqnarray}
      && \psi ^-({\mathsf s})=\langle ^-{\mathsf s}|\psi ^- \rangle =
         \langle ^-b,[j,{\mathsf s}]|\psi ^- \rangle \, , 
         \label{34a}\\
      && \phi ^+({\mathsf s})=\langle ^+{\mathsf s}|\phi ^+ \rangle =
         \langle ^+b,[j,{\mathsf s}]|\phi ^+ \rangle \, .
         \label{34b}
\end{eqnarray}
They have, according to our new hypothesis, the Hardy class property 
(\ref{20-b}) and (\ref{20b}).  $\phi ^+$ describes the prepared in-state
$(e^+e^-)$ and $|\phi ^+({\mathsf s})|^2=|\phi ^{in}({\mathsf s})|^2$ 
describes the
energy distribution of the beam.  $\psi ^-$ describes the observed
out-observable ($e^+e^-$, $\mu \bar{\mu}$, $\tau \bar{\tau}$,\ldots) which is
registered by the detector, and 
$|\psi ^-({\mathsf s})|^2=|\psi ^{out}({\mathsf s})|^2$
describes the detector efficiency.  Therefore they should be smooth, rapidly
decreasing functions.  In addition we require by (\ref{20-b}) and (\ref{20b})
certain analyticity properties for them which we conjectured 
in Section~\ref{sec:LShardycal} from the Lippmann-Schwinger equation (in the 
non-relativistic theory we attributed (\ref{20-b}) and (\ref{20b}) to a
causality principle~\cite{BOHM99}).

In order to be specific we shall consider the case that there are $N=2$ 
resonances in the $j$-th partial wave, each described by a first order pole 
at the position ${\mathsf s}={\mathsf s}_{R_1}$ and
${\mathsf s}={\mathsf s}_{R_2}$ in the second sheet.  The integrand in 
(\ref{33})
is thus analytic in the lower half second sheet except for the two 
poles at ${\mathsf s}={\mathsf s}_{R_i}$, and we can deform the contour of 
integration
in (\ref{33}) from the positive real line through the cut into the lower
half plane of the second sheet.  The r.h.s.\ of (\ref{33}) becomes (dropping
the $j$ notation)
\begin{eqnarray}
      (\psi ^-,\phi ^+)&=&\int_0^{-\infty _{II}}d{\mathsf s}\, 
       \langle \psi ^-|{\mathsf s}^-\rangle S_{II}({\mathsf s})
       \langle ^+{\mathsf s}|\phi ^+ \rangle \nonumber \\
      &&+\oint_{C_1} d{\mathsf s} \, 
       \langle \psi ^-|{\mathsf s}^-\rangle S_{II}({\mathsf s})
       \langle ^+{\mathsf s}|\phi ^+ \rangle \nonumber \\
      &&+\oint_{C_2} d{\mathsf s} \, 
       \langle \psi ^-|{\mathsf s}^-\rangle S_{II}({\mathsf s})
       \langle ^+{\mathsf s}|\phi ^+ \rangle \, ,
       \label{35}
\end{eqnarray}
where $C_i$ is the circle around the pole at ${\mathsf s}_{R_i}$, and the
first integral extends along the negative real axis in the second sheet 
(indicated by $-\infty _{II}$);  the integral along the infinite semicircle
is zero due to the assumed property of the integrand.  
The first term has nothing to do with any of the resonances, 
it is the non-resonant background term, 
\begin{equation}
       \int_{m_0^2}^{-\infty _{II}}d{\mathsf s}\, 
       \langle \psi ^-|{\mathsf s}^-\rangle S_{II}({\mathsf s})
       \langle ^+{\mathsf s}|\phi ^+ \rangle \equiv 
       \langle \psi ^-|\phi ^{bg}\rangle 
       \label{36}
\end{equation}
which we express as the 
matrix element of $\psi^-$ with a generalized vector $\phi^{bg}$
that is defined by it. It will be discussed further in Section~\ref{sec:tba}.

We now use the expansion around the pole ${\mathsf s}_{R_i}$
\begin{equation}
      S({\mathsf s})=\frac{R^{(i)}}{{\mathsf s}-{\mathsf s}_{R_i}}+R_0+
      R_1({\mathsf s}-{\mathsf s}_{R_i})+\cdots
      \label{37}
\end{equation}
for each of the two (or $N$) integrals {\it separately}.  The 
integrals around the poles, the pole terms, are calculated in the following
way:
\begin{eqnarray}
      (\psi ^-,\phi ^+)_{\rm pole \, term} &=&
      \oint_{\hookleftarrow C_i} d{\mathsf s}\, 
      \langle \psi ^-|{\mathsf s}^-\rangle S({\mathsf s})
       \langle ^+{\mathsf s}|\phi ^+ \rangle \label{38} \\
      &=&\oint_{\hookleftarrow C_i} d{\mathsf s}\, 
      \langle \psi ^-|{\mathsf s}^-\rangle 
      \frac{R^{(i)}}{{\mathsf s}-{\mathsf s}_{R_i}} 
       \langle ^+{\mathsf s}|\phi ^+ \rangle \label{39} \\ 
      &=&-2\pi i R^{(i)} \langle \psi ^-|{\mathsf s}_{R_i}^-\rangle 
          \langle ^+{\mathsf s}_{R_i}|\phi ^+ \rangle \label{40} \\
      &=&\int_{-\infty _{II}}^{\infty}d{\mathsf s} \,
       \langle \psi ^-|{\mathsf s}^-\rangle 
       \langle ^+{\mathsf s}|\phi ^+ \rangle 
       \frac{R^{(i)}}{{\mathsf s}-{\mathsf s}_{R_i}} \, . 
     \label{41} 
\end{eqnarray}
To get from (\ref{39}) to (\ref{40}), the Cauchy theorem has been applied; to
get from (\ref{39}) to (\ref{41}), the contour $C_i$ of each integral 
separately has been deformed into the integral along the real axis from
$-\infty _{II}<{\mathsf s}<+\infty$ (and an integral along the infinite 
semicircle,
which vanishes, because of the Hardy class property).  The
equality (\ref{40}) and (\ref{41}) is the Titchmarsh theorem for Hardy 
class functions.

The integral (\ref{41}) extends from ${\mathsf s}=-\infty _{II}$ in the second
sheet along the real axis to ${\mathsf s}=0$ and then from ${\mathsf s}=0$ to 
${\mathsf s}=+\infty$ in either sheet.  (It does not matter whether we take 
the second part of the integral over the physical values of $\mathsf s$, 
$m_0^2 \leq {\mathsf s}< \infty$, immediately below the real axis in the 
second sheet or in the first sheet immediately above the real axis).  The 
major contribution to the integral comes from the physical values
$m_0^2\leq {\mathsf s}<\infty$, if ${\mathsf s}_{R_i}$ is not too far from 
the real axis.  The integral in (\ref{41}) contains the Breit-Wigner amplitude 
\begin{equation}
       a_j^{BW_i}= \frac{R^{(i)}}{{\mathsf s}-{\mathsf s}_{R_i}} \, , \quad
       \text{but with} -\infty _{II}<{\mathsf s}<+\infty \, .
       \label{42}
\end{equation}
Unlike the conventional Breit-Wigner of (\ref{ajBW2}), for which $\mathsf s$
is taken (if one worries about these mathematical details) over 
$m_0^2 \leq {\mathsf s}<+\infty$, the Breit-Wigner (\ref{42}) is an idealized 
or exact Breit-Wigner whose domain extends to $-\infty _{II}$ in the second 
(unphysical) sheet.

By (\ref{41}) we have associated each resonance at ${\mathsf s}_{R_i}$ to an 
exact Breit-Wigner (\ref{42}) which we obtain by omitting the integral 
over the arbitrary function $\overline{\langle ^-{\mathsf s}|\psi ^- \rangle}
\langle ^+{\mathsf s}|\phi ^+\rangle\in {\cal S}\cap {\cal H}_-^2$ from 
(\ref{41}).  By (\ref{40}) we have associated each resonance at 
${\mathsf s}_{R_i}$ with vectors 
$|{\mathsf s}_{R_i}^-\rangle =|[j,{\mathsf s}_{R_i}],b^- \rangle$ which we call
Gamow kets.

We obtain a representation of the Gamow ket or Gamow vector by using the 
equality (\ref{40})=(\ref{41}) and omitting the arbitrary 
$\psi ^-\in {\mathbf \Phi} _+$
(which represents the decay products defined by the detector).  For this
defining relation of the relativistic Gamow kets or relativistic Gamow
vectors we shall use the notation that includes the degeneracy quantum numbers
$b$:
\begin{eqnarray}
      |[j,{\mathsf s}_{R_i}],b^- \rangle &=& 
       \frac{i}{2\pi}\int_{-\infty}^{\infty}d{\mathsf s} \, 
       |[j,{\mathsf s}],b^- \rangle \frac{1}{{\mathsf s}-{\mathsf s}_{R_i}} \, 
       \frac{\langle ^+{\mathsf s}|\phi ^+\rangle}
      {\langle ^+{\mathsf s}_{R_i}|\phi ^+\rangle}  \nonumber \\
      &=&\frac{i}{2\pi}\int_{-\infty}^{\infty}d{\mathsf s} \,
      |[j,{\mathsf s}],b^- \rangle \frac{1}{{\mathsf s}-{\mathsf s}_{R_i}} \, .
      \label{43}
\end{eqnarray}
The Gamow kets (\ref{43}) are a superposition of the exact---not 
asymptotically free~\cite{SHIRLINII}---``out states'' 
$|[j,{\mathsf s}],b^- \rangle$.  The degeneracy quantum numbers $b$ of the 
Gamow kets $|[j,{\mathsf s}_{R_i}],b^- \rangle$ are the same
as the ones chosen for the Dirac-Lippmann-Schwinger kets 
$|[j,{\mathsf s}],b^-\rangle$.  However, whereas for the 
Dirac-Lippmann-Schwinger
kets one can choose for $b=b_1,\ldots ,b_n$ the eigenvectors of any complete
set of observables, one does not have the same freedom for the $b$ in the Gamow
kets, since in the contour deformations
that one uses to get from (\ref{33}) to (\ref{35}) and ultimately to
(\ref{38})-(\ref{41}) one makes an analytic continuation in the variable
$\mathsf s$ to complex values.  If one chooses for $b$ quantum numbers that
also change when $\mathsf s$ is analytically continued, $b$ could not be kept
at one and the same value during this analytic continuation and the Gamow
vector on the l.h.s.\ of (\ref{43}) would be a complicated (continuous) 
superposition (integral) over different values of $b$ and not just a 
superposition over different values of $\mathsf s$.  For this reason, the 
momentum ${\bf p}$ is not a good choice for the quantum numbers $b$ in
(\ref{16}), because the momentum will also become complex if the energy
in the centre of mass rest frame becomes complex.  This is also the reason for 
which we choose the space components of the 4-velocity 
$\hat{\bf p}={\bf p}/\sqrt{\mathsf s}$ as the 
additional quantum numbers $b$, because then we can impose the condition that
${\bf p}$ will become complex in the analytic continuation in such a way 
that $\hat{p}^{\mu}=p^{\mu}/\sqrt{\mathsf s}$ will always be real.  This 
condition restricts the arbitrariness of the analytic continuation, it makes 
the momentum only ``minimally complex'' and keeps the representations of the
Lorentz subgroup of the Poincar\'e group ${\cal P}$ unitary. Only representations
of the space-time translations turn into (causal) semigroup 
representations.  
The homogeneous Lorentz transformations $U(\Lambda)$ are the same as in
Wigner's representations.
We will call this subclass of semigroup representations of ${\cal P}$ minimally
complex~\cite{RGVI-III,I-III}.

With (\ref{42}) and (\ref{43}) we have obtained for each resonance defined
by the pole of the $j$-th partial $S$-matrix at ${\mathsf s}={\mathsf s}_R$ an
``exact'' Breit-Wigner (\ref{42}) and associated to it a set of ``exact''
Gamow kets (\ref{43}).  These Gamow kets (\ref{43}) span, like the 
Dirac kets $|j, \sm],b\rangle$ in (\ref{16}), the space of an irreducible 
representation $[j,{\mathsf s}_R]$ of Poincar\'e transformations but, unlike 
the space spanned by the Dirac kets in (\ref{16}), 
the representation space spanned by the kets of (\ref{43}) is not 
the representation space of a unitary
group.  Thus we have the correspondence
\begin{equation}
      \begin{array}{lcl}
            \text{Exact Breit-Wigner} &\quad  \Longleftrightarrow \quad & 
           \text{Exact Gamow vectors} \\ [2ex]
            a_j^{BW_i}({\mathsf s})=\frac{R_i}{{\mathsf s}-{\mathsf s}_R} 
           &\quad \Longleftrightarrow \quad  & |[j,{\mathsf s}_R], f^-\rangle =
            \int d\mu (b)\, 
            |[j,{\mathsf s}_R],b^- \rangle f(b)  \\ [2ex]
           \text{for} \ -\infty _{II}<{\mathsf s}<+\infty  & \quad & 
           \text{for all} \ f(b) \in {\cal S}({\mathbb R} ^3), 
           -j\leq j_3\leq j \, .
      \end{array}
     \label{44}
\end{equation}
The Gamow vectors $|[j,{\mathsf s}_R],f^-\rangle$ have, according to (\ref{43})
and (\ref{16}), the representation
\begin{equation}
      |[j,{\mathsf s}_R],f^- \rangle =
      \frac{i}{2\pi}\int_{-\infty _{II}}^{+\infty}d{\mathsf s}\,
      |[j,{\mathsf s}],f^- \rangle \frac{1}{{\mathsf s}-{\mathsf s}_R} \, .
      \label{45}
\end{equation}
They are functionals on the Hardy space ${\mathbf \Phi} _+$, i.e., 
$|[j,{\mathsf s}_R],f^- \rangle \in {\mathbf \Phi} _+^{\times}$.  

Equation (\ref{45}) is reminiscent of the continuous basis vector expansion
(\ref{2.8a}) of $\psi ^- \in {\mathbf \Phi} _+\subset {\cal H}$ with respect to 
the generalized eigenvectors $|[j,{\mathsf s}],f^- \rangle$ of 
$P_{\mu}P^{\mu}$ with eigenvalue ${\mathsf s}$, where $\sm$ extends over 
$m_0^2\leq {\mathsf s}< \infty$.  However in (\ref{45}) the ``wave function''
$
\psi ^G({\mathsf s})\equiv\frac{i}{2\pi}\, \frac{1}{{\mathsf s}-{\mathsf s}_R}
$ 
is not a very well behaved, Hardy class wave function like 
$\psi ^-({\mathsf s})\in {\cal S}\cap {\cal H}_+^2$ of (\ref{2.8a}).  
Also in the exact 
Breit-Wigner ``wave function'' $\psi ^G({\mathsf s})$ in (\ref{45}), the 
variable ${\mathsf s}$ extends over $-\infty _{II}<{\mathsf s}<+\infty$.  Thus
the continuous linear superpositions (\ref{45}), which define the relativistic
Gamow vectors, are entirely different mathematical entities than the 
$\psi ^-$ of (\ref{2.8a}).  
The Gamow vectors $|[j,{\mathsf s}_R],f ^- \rangle$ and also the Gamow kets 
$|[j,{\mathsf s}_R],b^- \rangle$ of (\ref{43}) are functionals over the space
${\mathbf \Phi} _+$ (the $|[j,{\mathsf s}_R],b^- \rangle$ are in addition 
functionals over the Schwartz space, i.e., 
$\langle \psi ^-|[j,{\mathsf s}_R],b^- \rangle 
=f(b,j_3)\in {\cal S}({\mathbb R}^3)$ (for fixed value $\mathsf s$)).  
The equation (\ref{45}) and (\ref{43}) are functional equations over the space 
${\mathbf \Phi} _+$, and (\ref{43}) can be stated in terms of the smooth Hardy 
class functions 
$\overline{\psi}^-({\mathsf s})\equiv 
\langle \psi ^-|[j,{\mathsf s}],b^- \rangle \in {\cal S} \cap {\cal H}^2_-$ as
\begin{eqnarray}
   \langle \psi^-|[j,{\mathsf s_R}],b^-\rangle 
       &\equiv& -\frac{i}{2\pi} \oint d{\mathsf s}\,
         \langle \psi^-|[j,{\mathsf s}],b^-\rangle \,
        \frac{1}{{\mathsf s}-{\mathsf s}_R}
        \label{46} \\
   \langle \psi^-|[j,{\mathsf s_R}],b^-\rangle 
       &=&\frac{i}{2\pi} \int_{-\infty_{II}}^{+\infty} 
         d{\mathsf s}\, \langle \psi^-|[j,{\mathsf s}],b^-\rangle \, 
       \frac{1}{\mathsf{s}-\mathsf{s}_R}
         \, ,
      \label{47}
\end{eqnarray}
for all $\psi^-\in {\mathbf \Phi}_+$, and similarly for the vectors (\ref{45}).

The first equality (\ref{46}) is again the well known Cauchy formula for the 
analytic function 
$\overline{\psi}^- ({\mathsf s})=\langle \psi ^-|{\mathsf s}^-\rangle$.  
The second 
equality (\ref{47}) is the Titchmarsh theorem for the Hardy class function 
$\overline{\psi}^-({\mathsf s})$ in the lower half plane of the second 
sheet.  
The integration path extends as in (\ref{41}) along the 
real axis in the second sheet, which is only for physical values 
$m_0^2\leq {\mathsf s}<\infty$ the same as the integration along the 
real axis in the first sheet of (\ref{2.8b}).

The association (\ref{44}) of a space of 
vectors (\ref{45}) to the Breit-Wigner partial wave amplitude 
(\ref{42}) requires a very 
specific property of this amplitude, namely to be a Cauchy kernel, 
and the definition of the vectors (\ref{45}) and (\ref{43}) 
will not be possible 
for other arbitrary functions of $\mathsf s$ (e.g.\ not for the amplitude 
$a_j^{om}({\mathsf s})$ of (\ref{ajom1})).  Even for the Breit-Wigner
(\ref{ajBW2}) we had to extend the values of $\mathsf s$ from the 
phenomenologically testable values $m_0^2\leq {\mathsf s}<\infty$ in
(\ref{ajBW2}) to the negative axis and introduce an idealisation, the 
``exact'' Breit-Wigner (\ref{42}) for which $\mathsf s$ extends over 
$-\infty _{II}<{\mathsf s}<+\infty$.  
Only for the exact Breit-Wigner 
(\ref{42}) could we use the Titchmarsh theorem
in (\ref{41}), (\ref{47}) and associate to the 
amplitude $a_j^{BW}({\mathsf s})$ a vector which is defined by this exact 
Breit-Wigner amplitude.  And in order to apply the Titchmarsh theorem we had
to restrict the admissible wave functions $\overline{\psi}^-({\mathsf s})$ 
and $\phi ^+({\mathsf s})$ in (\ref{26a}), (\ref{32}) 
to be Hardy class in the lower half plane.  
That means we had to specify the in-state vector $\phi ^+$ and the 
out-observable vector $\psi ^-$ that can appear in the $S$-matrix 
element (\ref{32}) and (\ref{28}) to be in the spaces 
${\mathbf \Phi} _-$ and ${\mathbf \Phi} _+$, respectively.  
Only then could we 
define the Gamow kets $|[j,{\mathsf s}_R],b^- \rangle$ in terms of the 
Dirac-Lippmann-Schwinger kets $|[j,{\mathsf s}],b^-\rangle$ 
by e.g. (\ref{47}) as generalized vectors
or functionals over the Hardy class space ${\mathbf \Phi} _+$.
The Gamow vectors cannot even be defined as functionals
over the Schwartz space ${\mathbf \Phi}$ like the usual Dirac kets. 
(Similarly we can define another kind of Gamow ket 
$|[j,{\mathsf s}_R^*],b ^+\rangle \in {\mathbf \Phi} _-^{\times}$
in terms of the Dirac-Lippmann-Schwinger kets $|[j,{\mathsf s}],b^+ \rangle$
for the resonance pole at ${\mathsf s}_R^*=(M_R+i\Gamma /2)^2$ in the upper 
half plane of the second sheet).  Thus the Hardy spaces ${\mathbf \Phi} _-$ 
and ${\mathbf \Phi} _+$, and therewith the new hypothesis of 
Section~\ref{sec:LShardycal}, had to be introduced (as in the non-relativistic 
theory~\cite{JPM81}) in order to be able to construct vectors
(\ref{43}), (\ref{45}) and (\ref{46})
with a Breit-Wigner energy distribution.

From these Gamow vectors we can now calculate consequences without any 
further mathematical assumption.  These predictions are that (\ref{45}) and/or
(\ref{43}) are generalized eigenvectors of the operators 
\begin{equation}
      P_{\mu}P^{\mu}\, , \ P^0=H=H_0+V \, , \ e^{iHt} (e^{-iH^{\times}t}) \, .
      \label{49}
\end{equation}
We shall now show this for one example $P_\mu P^\mu$.
We consider the vector $\psi^{'-}=P_\mu P^\mu\psi^-$; this makes
sense because $\psi^{'-}\in{\mathbf\Phi}_+$ for any 
$\psi^-\in{\mathbf\Phi}_+$, because ${\mathbf\Phi}_+$ 
is constructed such that all observables are continuous operators 
with respect to the topology in ${\mathbf\Phi}_+$ (explicitly given by the
space ${\cal S}$ in (\ref{20-b})). 
We now use (\ref{47}) for $\psi^{'-}=P_\mu P^\mu\psi^-$:
\begin{eqnarray}
      \langle P_\mu P^\mu\psi^-|[j,\sm_R], b^-\rangle
     &=&\langle \psi^-|(P_\mu P^\mu)^{\times}[j,{\mathsf s}_R],b^-\rangle 
                                                        \nonumber\\
      &=&\frac{i}{2\pi}\int_{-\infty_{II}}^{+\infty} 
      d{\mathsf s}\, \langle P_\mu P^\mu\psi^-|[j,{\mathsf s}],b^-\rangle
      \frac{1}{{\mathsf s}-{\mathsf s}_R}\nonumber \\
      &=&\frac{i}{2\pi}\int_{-\infty_{II}}^{+\infty} 
      d{\mathsf s}\, 
      \langle \psi^-|(P_\mu P^\mu)^{\times}[j,{\mathsf s}],b^-\rangle
       \, \frac{1}{{\mathsf s}-{\mathsf s}_R}\nonumber\\
      &=&\frac{i}{2\pi}\int_{-\infty_{II}}^{+\infty} 
      d{\mathsf s}\, \langle \psi^-|[j,{\mathsf s}],b^-\rangle 
      \,\, \sm \,\, \frac{1}{{\mathsf s}-{\mathsf s}_R} 
        \nonumber\\ 
      &=&\sm_R\langle \psi^-|[j,{\mathsf s}_R],b^-\rangle \,.
      \label{3-32}
\end{eqnarray}
The next to the last step makes use of (\ref{7a}) and the last step is again 
the Titchmarsh theorem as in (\ref{47}) but this time 
for the function 
$
\overline{\psi} '^{-}({\mathsf s})\equiv\overline{\psi}^-({\mathsf s})\,\sm
$, 
which is 
also a Hardy class function  
$\overline{\psi}^{-}\,({\sm})\,\sm\in{\cal S}\cap{\cal H}^2_-$ if
$\overline{\psi}^{-}\,({\sm})\in{\cal S}\cap{\cal H}^2_-$. 
We write (\ref{3-32})
as a functional equation over the space ${\mathbf \Phi} _+$ omitting the 
arbitrary $\psi ^- \in {\mathbf \Phi} _+$,
\begin{equation}
      (P^\mu P_\mu)^{\times} |[j,{\mathsf s}_R],b,j_3 ^-\rangle
      ={\mathsf s}_R \, |[j,{\mathsf s}_R],b,j_3 ^-\rangle 
\label{51} 
\end{equation}
Similarly, we find from (\ref{7a}) for the full Hamiltonian $P^0=H=H_0+V$
at rest:
\begin{equation}
       H^{\times}|[j,{\mathsf s}_R],b_{\text{rest}},j_3 ^-\rangle=
      \sqrt{{\mathsf s}_R} \, |[j,{\mathsf s}_R], b_{\text{rest}},j_3 ^-\rangle
       \label{50}
\end{equation}
and
\begin{equation} 
     (P_\mu P^\mu)^{\times} |[j,{\mathsf s}_R], f^-\rangle
      ={\mathsf s}_R \, |[j,{\mathsf s}_R], f^-\rangle \, , \quad 
      \text{for any} \ f(j_3,b)\in {\cal S}({\mathbb R} ^3)\, ,
\label{52}
\end{equation}
i.e.\ for the whole representations space of $[j,\sm_R]$ of (\ref{44}).
In here ${\mathsf s}_R$ is the resonance pole position.  
These equations state 
that the Gamow kets and the whole space of vectors (\ref{44}) spanned by 
the Gamow kets are generalized eigenvectors of the self-adjoint invariant
mass square operator with complex eigenvalue ${\mathsf s}_R$.  If we use the 
parameterisations (\ref{5a})-(\ref{5e}), then the 
eigenvalue of $H$ can be written as
\begin{equation}
      {\rm eigenv}(H)=\sqrt{{\mathsf s}_R}=(M_R-i\frac{\Gamma _R}{2})
      =\sqrt{\bar M_Z^2-i\bar M_Z^2\bar \Gamma_Z}
      =\sqrt{\frac{(m^2_1 -im_1\Gamma_1)}{(1+\frac{\Gamma_1}{m_1})}}\, .
      \label{53}
\end{equation}
 
The results (\ref{3-32})-(\ref{52}) do not yet distinguish any of the 
parameterisations of ${\mathsf s}_R$.  In order to 
address the question of whether there is any preferred physical significance
to one of the parameterisations in (\ref{53}), we shall consider the 
Poincar\'e transformations $(a,\Lambda )$, where $a$ is the 4-vector of 
space time translations and $\Lambda$ is a $4\times 4$ Lorentz matrix.  
We shall give here only the action of time translations 
$(a=(t,0,0,0), \Lambda ={\bf 1})$ on the basis vectors at rest
$|[j,\sm_R],b_{\text{rest}}^-\rangle$.  
For the general transformation property of the Poincar\'e semigroup and
a detailed proof of the semigroup property we refer to the forthcoming 
\cite{I-III}.

We know that in the stable particle representations $[j,m^2]$ the
time translation is represented by the operator 
$U((t, {\bf 0}),{\bf 1})\equiv U(t)=e^{iHt}$ 
in the Hilbert space ${\cal H}(j,m^2)$ and therefore also in the 
space ${\cal H}$ of (\ref{21-}). 
The time translation in the subspace ${\mathbf\Phi}_+$ will therefore be the 
restriction $U_+(t)=U(t)|_{{\mathbf\Phi}_+}=e^{iHt}|_{{\mathbf\Phi}_+}$ 
to this subspace.
The time translation in the space ${\mathbf\Phi}^\times_+$ will be 
the extension of the unitary group operator 
$U^\dagger (t)=e^{-iHt}\subset U^\times_+(t)$, whenever it can be defined.
In order that $(e^{iHt})^\times =U^\times_+(t)$ can be defined the 
operator $U_+(t)$ needs to be a continuous operator with respect to the 
topology in ${\mathbf\Phi}_+$ (cf.~Appendix). 
Therefore the question is: for which value of the parameter $t$ is
$U_+(t)$ (and therewith also $U_+^\times(t)$) 
a continuous operator in ${\mathbf\Phi}_+$,
\begin{equation}
   {\mathbf \Phi}_+ \ni \psi^- \to U_+(t)\psi^-\equiv\psi^{'-} 
          \in {\mathbf\Phi}_+ \,.
\label{3-37a}
\end{equation}
We show below that (\ref{3-37a}) holds only for $t\geq 0$ so that
also $U_+^\times(t)$ can only be defined for $t\geq 0$.
For those values of $t$ for which $U_+(t)$ and $U_+^\times(t)$ 
is defined, we have for all
$m^2_0\leq \sm < \infty$ and then also for all real
$\sm$\footnote{Using
the van Winter theorem for Hardy class functions, cf.~Appendix A2 of 
\cite{NONRELAa}.}:
\begin{eqnarray}
    \langle \psi^{'-}|[j,\sm],b_{\text{rest}}^-\rangle  
  &=& \langle U_+(t) \psi^{-}|[j,\sm],b_{\text{rest}}^-\rangle \nonumber \\
  &=& \langle\psi^-|U^\times_+(t)|[j,\sm], b_{\text{rest}}^-\rangle \nonumber \\ 
  &=& \langle e^{iHt}\psi^-|[j,\sm],b_{\text{rest}}^-\rangle \nonumber \\
  &=& e^{-i\sqrt{\sm}\,t}\langle \psi^{-}|[j,\sm],b_{\text{rest}}^-\rangle  
      \,\,{\rm for\,\, all}\,\,\psi^-\in{\mathbf\Phi}_+
\label{3-37b}
\end{eqnarray}
Omitting the arbitrary $\psi^-\in{\mathbf\Phi_+}$ we write this
as a functional equation:
\begin{equation}
      e^{-iH^{\times}_+ t}|[j,{\mathsf s}],b_{\text{rest}}^-\rangle\,=
        e^{-i{\sqrt{\mathsf s}}\,t}
          |[j,{\mathsf s}],b_{\text{rest}}^-\rangle\, 
           , \,\,\,{\rm for}\,\,\, t\geq 0 \,\,\,{\rm only},
\label{54}
\end{equation}
where 
\begin{equation}
e^{-iH^\times_+ t}\equiv U^\times_+(t)=(e^{iHt})^\times\,,\,\,\,\,t\geq 0\,.
\label{3-37d}
\end{equation}
The first term in (\ref{3-37d}) is so far only a definition, but 
one can show that $H^\times_+ \supset H^\dagger$, the
conjugate operator of $H|_{{\mathbf\Phi}_+}\equiv H_+$, 
and extension of the operator
$H^\dagger = H$ to ${\mathbf\Phi}_+^\times$ (cf.~Appendix) is indeed the 
generator of the semigroup $U^\times_+(t)$ \cite{Sujeewa}. 

The solutions of the other Lippmann-Schwinger
equation with $+i\epsilon$, $|[j,\sm],b^+\rangle\in{\mathbf\Phi}^\times_-$, 
have a time evolution given by the other semigroup $U^\times_-(t)$ with
$t\leq 0$,
\begin{equation}
   e^{-iH^\times_- t}|[j,\sm], b_{\text{rest}}^+\rangle 
       = e^{-i\sqrt{\sm}t}|[j,\sm],b_{\text{rest}}^+\rangle\,\,\,\,
             {\rm for}\,\,t\leq 0\,\,{\rm only}.
\label{3-41}
\end{equation}
In here $U^\times_-(t)=e^{-iH^\times_-t}$ 
(and $H^\times_-$) are analogously
defined in the RHS (\ref{2-18}) as the extension of the $U^\dagger(t)$ 
(and of $H^\dagger$) to the space ${\mathbf\Phi}^\times_-$.

$\lbrace$Brief justification of the semigroup condition $t\geq 0$:
The complex conjugate of (\ref{3-37b}) can be written in the notation
of (\ref{2-9a}), (\ref{20-b})
\begin{equation}
\psi^{'-}(\sm)=e^{i\sqrt{\sm}\, t}\psi^-(\sm)
\label{3-37e}
\end{equation}
The question: For which $t$ in (\ref{3-37a}) is $\psi^{'-}\in{\mathbf\Phi}_+$
if $\psi^{-}\in{\mathbf\Phi}_+$, can thus be formulated: for which $t$ 
in (\ref{3-37e})
is $\psi^{'-}(\sm)\in{\cal S}\cap {\cal H}^2_+$ if 
$\psi^{-}(\sm)\in{\cal S}\cap{\cal H}^2_+$. 
It is easy to see and intuitively clear that only for $t\geq 0$ will
$e^{i\sqrt{\sm}\, t}\psi^-(\sm)$ decrease sufficiently fast in the 
upper half plane so that $\psi^{'-}(\sm)$ is also Hardy class. 
For the detailed proof of the semigroup property for the general 
Poincar\'e transformation we refer
to \cite{I-III,Sujeewa}.$\rbrace$ 

We obtain now the action of 
$e^{-iH^{\times}t}$ on the Gamow ket at rest using 
(\ref{47}) for the vector $e^{iHt}\psi^-\equiv\psi '^- (t)$, where 
$t$ is the time in the rest frame of the resonance:
\begin{eqnarray}
      \langle\psi^-|e^{-iH^{\times}t}|[j,{\mathsf s}_R], b_{\text{rest}}^-\rangle
     &=&\langle e^{iHt}\psi^-|[j,{\mathsf s}_R],
     b_{\text{rest}}^-\rangle \nonumber\\
      &=&\frac{i}{2\pi}\int_{-\infty_{II}}^{+\infty} 
      d{\mathsf s}\, \langle e^{iHt}\psi^-|[j,{\mathsf s}],
      b_{\text{rest}}^-\rangle
      \frac{1}{{\mathsf s}-{\mathsf s}_R}\nonumber \\
      &=&\frac{i}{2\pi}\int_{-\infty_{II}}^{+\infty} 
      d{\mathsf s}\, 
      \langle \psi^-|e^{-iH^{\times}t}|[j,{\mathsf s}],b_{\text{rest}}^-\rangle
       \, \frac{1}{{\mathsf s}-{\mathsf s}_R}\nonumber\\
      &=&\frac{i}{2\pi}\int_{-\infty_{II}}^{+\infty} 
      d{\mathsf s}\, \langle \psi^-|[j,{\mathsf s}],b_{\text{rest}}^-\rangle 
      e^{-i\sqrt{{\mathsf s}}t} \, \frac{1}{{\mathsf s}-{\mathsf s}_R} 
        \nonumber\\ 
      &=&e^{-i\sqrt{{\mathsf s}_R}t}
       \langle \psi^-|[j,{\mathsf s}_R],b_{\text{rest}}^-\rangle \, ,
       \quad{\rm for}\ t\geq 0 \, .
      \label{55}
\end{eqnarray}
The next to the last step makes use of (\ref{54}) 
and the last step is again 
the Titchmarsh theorem as in (\ref{47}) or (\ref{3-32}) but this time 
for the function 
$
\overline{\psi} '^{-}({\mathsf s})
\equiv\overline{\psi}^-({\mathsf s})e^{-i\sqrt{{\sm}}t}
\in {\cal S}\cap {\cal H}^2_-
$, 
which is also a Hardy class function if $\overline{\psi}^-(\sm)$ is; however 
$\psi '^{-}({\mathsf s})$ is
Hardy class only if $t\geq 0$.    
For $t<0$ the time evolution operator $e^{-iH^{\times}t}$ on 
$|[j,{\mathsf s}],b^-\rangle$ and on
$|[j,{\mathsf s}_R],b^-\rangle$ does not exist.  

We write (\ref{55})
as a functional equation over the space ${\mathbf \Phi} _+$ omitting the 
arbitrary $\psi ^- \in {\mathbf \Phi} _+$,
\begin{equation}
      |\psi^G_{j,{\mathsf s}_R}(t)^-\rangle \equiv 
      e^{-iH^{\times}t} |[j,{\mathsf s}_R],b_{\text{rest}}^-\rangle 
     =e^{-iM_Rt}e^{-\frac{\Gamma_Rt}{2}} |[j,{\mathsf s}_R],
      b_{\text{rest}}^-\rangle 
      \, , \quad  t\geq 0 \, . 
      \label{56}
\end{equation}
In (\ref{56}) we have used the most suitable parameterisation 
for $\sqrt{\sm_R}$ of which a few other popular parametrisations 
are also given in (\ref{53}).  

The result (\ref{56}) is a mathematical consequence of the 
new hypothesis of (\ref{2-21}). It has two physical consequences:
\begin{enumerate}
    \item   The time evolution of the Gamow vectors has a preferred direction
                of time, i.e., it is not reversible.  
    \item   The time evolution of the relativistic Gamow vectors is 
                exponential with the decay constant (inverse lifetime)
                $\Gamma_R$.
\end{enumerate}

The time asymmetry, $t \geq 0$, has been 
discussed in detail elsewhere \cite{BOHM99} and also in \cite{TIASYa,TIASYb}
and is not the main interest of this paper.
Though the 
``profoundly irreversible'' character of a quantum decay has
been mentioned in the past~\cite{IRRQD}, it usually has been attributed to 
other principles (effect of the environment, collapse of the wave function, 
decoherence), probably because of the misconception that time evolution in 
quantum mechanics must be given by the reversible unitary group as is 
dictated by the mathematics of the Hilbert space.  
The irreversibility on the microphysical 
level expressed by the time asymmetry (\ref{56}) was 
the first unintended and the most surprising result of our
time asymmetric quantum theory \cite{JPM81}.  
It is a consequence of the time asymmetric 
boundary conditions (\ref{2-21}) and has nothing to do with violation of
time reversal invariance (which is a statement about the $T$-transformation
property of the Hamiltonian $H$) and it is not known to us whether it can be 
connected to entropy increase \cite{Gellmann}.  

In retrospect this
time asymmetry is no longer shocking, since Maxwell's theory and general 
relativity theory also use time asymmetric boundary conditions 
for time symmetric dynamical equations. 
Time asymmetric (purely outgoing) boundary
conditions in place of the Hilbert space assumption have also been suggested
in the past for quantum mechanics~\cite{29}. 
The germ of this  
time asymmetry is already inherent in the 
Lippmann-Schwinger equations \cite{L-Sch} and in the Feynman rules, 
by the infinitesimal
$i\epsilon$~\cite{feynman}. 
But surprisingly, none of those papers arrived at a semigroup
evolution like (\ref{54}) and (\ref{3-41}).
This semigroup evolution of the in- and out-planewave states is a 
manifestation of a fundamental time asymmetry in quantum 
scattering -- independently of whether resonance formation is involved
or not.

It can be shown that the time evolution semigroup (\ref{56}) is the rest 
frame version of a causal semigroup $P_+=(a_+,\Lambda)$ of Poincar\'e 
transformations where $\Lambda$ is a proper orthochronous Lorentz 
transformation and $a_+=(a_0,{\bf a})$ is a 4-vector that fulfills 
$b\cdot a=\sqrt{1+{\bf b}^2}\, a_0-{\bf b} \cdot {\bf a}\geq 0$ for any 
${\bf b}\in {\mathbb R}^3$.  It means that the relativistic Gamow vectors can 
only undergo Poincar\'e transformations into the forward light 
cone~\cite{RGVI-III,I-III}.

The second consequence of (\ref{56}), the exponential time evolution 
of the Gamow vectors, is not a surprize, because it was the 
intended property for which the Gamow vectors of the non-relativistic 
theory were originally constructed.  
However that the exponential decay constant is $\Gamma_R$ and not
any of the other $\Gamma$'s of (\ref{53}) is a new prediction
of the relativistic Gamow vector.
In the non-relativistic theory \cite{NONRELAa,NONRELAb,NONRELAc}
the exponential time evolution of the non-relativistic Gamow vectors,
\begin{equation}
      \psi ^G(t)\equiv e^{-iH^{\times}t}\psi ^G=
      e^{-iE_Rt}e^{-\Gamma_R /2 t}\psi ^G \,,
      \label{3.41b}
\end{equation}
was the property needed to calculate from the Born probabilities 
${\cal P}_{\eta}(t)$ (\ref{25}) the exponential law for the 
partial decay rates (\ref{26a})
\begin{equation}
      \dot P_{\eta} (t)
      =e^{-\Gamma_R t} \Gamma _{\eta} \, ,
      \label{3.45}
\end{equation}
and the exact Golden Rule 
\begin{equation}
       \Gamma _{\eta} =2\pi \sum_{b=b_\eta} \int_{0}^{\infty} d{E} \, 
      |\langle b,E|V|\psi^G\rangle|^2 
      \, \frac{\frac{\Gamma _R}{2\pi}}{(E-E_R)^2+
      (\frac{\Gamma _R}{2})^2}
     \label{3.46}
\end{equation} 
for the partial initial decay rate (also called partial width for the decay 
into the channel $\eta$).  For $t\to 0$ and in the Born approximation, 
defined by
\begin{equation}
     \psi^G\to f^D \, ,  \ \frac{\Gamma_R}{M_R}\to 0 \, , \ 
    \text{where}\ H_0f^D=E_Df^D \, ,\, H_0=H-V \,,
     \label{3.47} 
\end{equation}
one obtains from the ``exact Golden Rule'' (\ref{3.46}) the
Fermi-Dirac's Golden Rule, for the initial decay rate
\begin{equation}
      \Gamma _{\eta}
      =2\pi \sum_{b=b_\eta}   \int_{0}^{\infty} dE \,
      |\langle b ,E|V|f^D\rangle |^2\delta (E-E_D) \, .
      \label{3.48}
\end{equation}
The sum in (\ref{3.46}) and (\ref{3.48}) extends over all values of the quantum
numbers $b$ which characterize the channel $\eta$.  
If one also sums 
over all channels $\eta$ (using the condition 
${\cal P}(0)\equiv \sum_\eta {\cal P}_{\eta}(0)=0$, which means 
the probability to find any decay product $\eta$ at $t=0$, is zero
\cite{NONRELAb}), one obtains:
\begin{equation}
       \sum_{\eta} \Gamma _{\eta} \equiv \sum_{\eta} \dot P_{\eta}(0)
      =\Gamma_R \, .
      \label{3.49}
\end{equation}
This means that the sum over all partial initial decay rates is the width
of the Breit-Wigner, which according to the exponential law (\ref{3.45}) is 
also the inverse lifetime.

The same result one expects also for the relativistic 
case, this means
from the exponential time evolution (\ref{56}) of the
relativistic Gamow vectors at rest,
follows the exponential decay law (\ref{3.45}) with
$t$ being the 
time in the rest frame of the decaying state
$R$, (i.e., for 
$b_{\text{rest}}\equiv \hat{\bf p}_{\text{rest}}={\bf 0}$ and therefore
${\bf p}_{\text{rest}}=\sqrt{{\mathsf s}_R} \hat{\bf p}_{rest}=
{\bf 0}$).  
Therefore we conclude from the results (\ref{55}), (\ref{56}) that 
\begin{equation}
       \Gamma _R=\frac{1}{\tau} \, ,
       \label{3.52}
\end{equation}
where $\tau$ is the lifetime of the relativistic resonance $R$ (average
lifetime in the rest frame).

The parameter lifetime $\tau$ of a relativistic decaying state is 
measured by the counting rate (\ref{24}) as a function of time $t$ in the rest
frame or (practically always) as a function of the distance 
$d=\hat{\bf p}t=\gamma {\bf v}t$ which the quasistable particle with 
space components of the 4-velocity 
$\hat{\bf p}=\gamma {\bf v}= (1-v^2)^{-1/2} {\bf v}=\gamma {\mathbf \beta}c$
travels in the lab frame:
\begin{equation}
      \frac{\Delta N_{\eta}(t)}{\Delta t}=
      \dot N _{\eta}(0)e^{-t/\tau}= e^{-d\frac{1}{\hat{p}}\frac{1}{\tau}}=
      e^{-d\frac{1}{\gamma |{\bf v}|}\frac{1}{\tau}}=
      e^{-\frac{d}{\gamma \beta c}\frac{1}{\tau}} \, .
      \label{3.53}
\end{equation}
For the $Z$-boson (and for hadron resonances) $\hbar /\Gamma$ is too
small to be measured in this way and one can only measure $\Gamma$ from 
the lineshape (\ref{ajom1}).  For other weakly decaying particles 
\cite{Kaon}, like the $K^0$, one measures the lifetime 
$\tau$ using (\ref{3.53}) but one
cannot resolve the lineshape (\ref{ajom1}) because $\hbar /\tau$ is too
small.  Thus there may be no way to test the relation (\ref{3.52}) ever. 

The width-lifetime relation (\ref{3.52}) is a result for the relativistic
Gamow vectors in the same way as (\ref{22}) is a result for the 
non-relativistic Gamow vectors.  Without Gamow vectors neither
(\ref{3.45}), (\ref{3.46}) nor (\ref{3.52}) can be proven as
an exact relation.  The difference between the relativistic and 
non-relativistic case is that in the latter there was never any doubt 
regarding the meaning of the width, it is the full width at half-maximum
of the non-relativistic Breit-Wigner (\ref{23}).  For the relativistic
resonance one was not sure which part of the $j$-th partial wave 
amplitude (\ref{31}) one should assign to the resonance per se, i.e., whether
(\ref{ajom1}) or (\ref{ajBW2}) or any other part of $a_j({\mathsf s})$ 
describes
the resonance~\cite{SIRLIN,WILLENBROCK,STUART}.  Neither was one sure, that 
even when (\ref{ajBW2}) was chosen (using the $S$-matrix pole definition), 
which of the parameterisations (\ref{5a}), (\ref{5b}), (\ref{5d}) should be 
used to define
the ``width'' and the real mass of the resonance.  
The result 
(\ref{56}) for the relativistic Gamow vector fixes this: the mass is
$M_R={\rm Re}(\sqrt{{\mathsf s}_R})$ (the coefficient of the phase 
for the time evolution (\ref{56})) and the width is 
$\Gamma _R=-2{\rm Im}(\sqrt{{\mathsf s}_R})$
(the coefficient of the exponential decay).  Then (\ref{3.52}) 
holds universally.  This means that if for every resonance and every decaying 
state one wants (\ref{22}) to hold independently of whether both are 
measurable, then the 
relativistic Gamow vectors predict that $M_R$ is the mass and $\Gamma _R$
is the ``width'' of the relativistic quasistable particle or resonance.  
For the $Z$-boson, this means that neither $(\bar{M}_Z,\bar{\Gamma}_Z)$ 
nor $(m_1=M_Z,\Gamma _1=\Gamma _Z)$, as quoted in the Particle Data 
Table~\cite{CASO}, is the 
$Z$-boson mass and width but the third possibility (\ref{MRvalue})
of Section~\ref{sec:introduction}.

\section{From a single Breit-Wigner to their superpositions}
\label{sec:tba}

The exact Breit-Wigner amplitude $a_j^{BW_i}(\mathsf s)$ 
of (\ref{42}) was one particular part of the $j$-th 
partial $S$-matrix (\ref{37}) namely the one 
that was obtained from the pole of $S_j({\mathsf s})$ 
and which we considered separately in Section~\ref{sec:fSMpoGvec}.  With 
our mathematical hypothesis of Section~\ref{sec:LShardycal}, it was 
completely natural to treat each integral around a resonance pole at
${\mathsf s}_{R_i}$ separately and assign to each 
a Breit-Wigner amplitude (\ref{42}) and a corresponding Gamow 
vector (\ref{44}). In an experiment it is not that easy to
separate the resonance term(s) from the remainder 
of the $S$-matrix, since the $j$-th partial cross section does not
only contain the resonance part but contains 
also the non-resonant background of the 
scattering process in the amplitude $a_j({\mathsf s})$.  We now want to 
consider 
the whole $S$-matrix element (\ref{35}).  Inserting (\ref{31}) into (\ref{35}) 
and using the definition (\ref{36}) we can write the $j$-th partial $S$-matrix 
element $(\psi^-,\phi^+)$ as a discrete 
sum over Gamow vectors and the background term,
\begin{equation}
       (\psi ^-,\phi ^+)=\langle \psi ^-|\phi ^{bg}\rangle +
       \sum_i \langle \psi ^-|{\mathsf s}_{R_i}^-\rangle (2\pi /i)
       \langle ^+{\mathsf s}_{R_i}|\phi ^+\rangle \, .
       \label{4.1}
\end{equation}
Omitting the arbitrary $\psi ^-\in {\mathbf \Phi}_+$ (the observable) one 
writes Eq.~(\ref{4.1}) as a functional equation in the space 
${\mathbf \Phi}_+^{\times}$ and obtains the following expansion of the prepared 
in-state $\phi ^+ \in {\mathbf \Phi} _-$:
\begin{equation}
       \phi ^+=\phi ^{bg}+\sum_i|{\mathsf s}_{R_i}^- \rangle c_{R_i} \, ,
      \quad \text{where}\ c_{R_i}= (2\pi /i)
       \langle ^+{\mathsf s}_{R_i}|\phi ^+\rangle \, .
      \label{4.2}
\end{equation}
In this way the in-state $\phi ^+$ has been decomposed into a vector 
representing the non-resonant part $\phi ^{bg}$ and a sum over the Gamow
vectors representing resonance states.  The complex eigenvalue resolution
(\ref{4.2}) is an alternative generalized eigenvector expansion 
to Dirac's eigenvector
expansion (\ref{17}),
\begin{equation}
     {\mathbf\Phi}_- \ni \phi ^+=\int_0^{\infty}d{\mathsf s} \, 
          |{\mathsf s}^+\rangle 
        \langle ^+{\mathsf s}|\phi ^+ \rangle \,\, ,\,\,
           |\sm^+\rangle \in {\mathbf \Phi}^\times_-\, .
       \label{4.3}
\end{equation}
While Eq.~(\ref{4.3}) expresses the in-state $\phi ^+$ in terms of 
the Lippmann-Schwinger kets 
$|{\mathsf s}^+\rangle\in {\mathbf\Phi}^\times_-$, 
which are generalized 
eigenvectors of the Hamiltonian $H$ with real 
eigenvalue $\sqrt{\mathsf s}$, Eq.~(\ref{4.2}) is an expansion of
$\phi^+ \in {\mathbf\Phi}^\times_+$
in terms of eigenkets 
$|{\mathsf s}_{R_i} ^-\rangle \in {\mathbf\Phi}^\times_+$ of 
the same {\it self-adjoint}\/ Hamiltonian $H$ with 
complex generalized eigenvalue 
$\sqrt{{\mathsf s}_{R_i}}=M_{R_i}-i\Gamma_{R_i}/2$ and the 
$\phi^{bg} \in {\mathbf\Phi}^\times_+$.

The term $\phi^{bg}$ is defined by (\ref{36})
and is therefore an element of ${\mathbf \Phi}^{\times}_+$.  We want to 
rewrite (\ref{36}) into a more familiar form.  According to the van Winter 
theorem~\cite{VANWINTER}, a Hardy 
class function on the negative real axis is uniquely
determined by its values on the real positive axis (cf.~Appendix A2 
of~\cite{NONRELAa}).  Therefore one can use the Mellin 
transform to rewrite the integral on the l.h.s.~of (\ref{36})
into an integral over the interval $m^2_0 \leq {\mathsf s} < \infty$ and obtain
\begin{equation}
      \langle \psi^-|\phi^{bg} \rangle 
      = \int_{m^2_0}^{-\infty_{II}} d{\mathsf s} \, 
         \langle \psi^-|{\mathsf s}^- \rangle S_j({\mathsf s})
         \langle ^+ {\mathsf s}|\phi^+ \rangle
      = \int_{m^2_0}^{\infty} d{\mathsf s} \, 
        \langle \psi^-|{\mathsf s}^- \rangle b_j({\mathsf s})
       \langle ^+{\mathsf s}|\phi^+ \rangle \, ,
       \label{4.4}
\end{equation}
where $b_j({\mathsf s})$ is uniquely defined by the values of 
$S_j({\mathsf s})$ on the negative real axis.  Without more specific
information about $S_j({\mathsf s})$, we cannot be
certain about the energy dependence of the background
$b_j({\mathsf s})$.  If there are no further poles or singularities 
besides those included in the sum, then $b_j({\mathsf s})$ 
is likely to be a slowly varying function of 
$\mathsf s$~\cite{GADELLA2}.  Omitting the arbitrary 
$\psi ^- \in {\mathbf \Phi} _+$
we write the expansion for the non-resonant background part $\phi ^{bg}$ of the
prepared in-state vector $\phi ^+$ as:
\begin{equation}
      |\phi^{bg} \rangle = \int_{m^2_0}^{\infty} d{\mathsf s} \,
      |{\mathsf s} ^-\rangle \langle ^+ {\mathsf s}|\phi^+ \rangle 
     b_j(\mathsf s) \, ,
    \label{4.5}
\end{equation}
Inserting (\ref{4.5}) into (\ref{4.2}) we obtain the basis vector expansion of every 
$\phi^+\in {\mathbf \Phi} _-$, 
\begin{equation}
      \phi^+ =\sum_{i} |{\mathsf s}_{R_i} ^-\rangle c_{R_i}
         +\int_{m^2_0}^{\infty} d{\mathsf s} \, 
          |{\mathsf s} ^-\rangle \langle^+ {\mathsf s}|\phi^+ \rangle 
         b_j(\mathsf{s}) \,\,\, ; \,\,\, 
           |\sm_{R_i} ^-\rangle\, , \, |\sm ^-\rangle 
                \in {\mathbf\Phi}_+^\times
       \label{4.5a}
\end{equation}
(We have assumed as in (\ref{4.3}) that there are no bound states of 
$H$.  Otherwise one would have in addition to the r.h.s.~of (\ref{4.3}) and
(\ref{4.5a}) the discrete sum over the bound states, which are orthogonal to
the rest).  

The vector (\ref{4.5}) does not have an exponential time evolution,
but all the Gamow vectors $|\sm_{R_i}^-\rangle$ in the 
basis vector expansion (\ref{4.2}) evolve exponentially.  
The much debated deviation from the exponential decay
law~\cite{KHALFIN} has its origin in this background integral (\ref{4.5}) for 
the prepared in-state $\phi ^+$.  The experimental ingenuity in 
establishing the exponential decay law for each Gamow state
is to suppress or exclude as much as possible this background 
and the effect of the other interfering Gamow vectors in the
analysis of the experimental data \cite{Kaon}.  

We now rewrite (\ref{35}) in a different form.  In place of
$(\psi ^-, \phi ^+)$ on the l.h.s.~of (\ref{35}), we write the r.h.s.~of
(\ref{33}), and on the r.h.s.~of (\ref{35}) we use (\ref{41}) and 
(\ref{4.4}).  Then we obtain
\begin{equation}
     \int_{m_0^2}^{\infty}d{\mathsf s}\, 
      \langle \psi ^-|{\mathsf s}^-\rangle 
      S_j({\mathsf s}) \langle ^+{\mathsf s}|\phi ^+\rangle =
     \int_{m_0^2}^{\infty}d{\mathsf s}\, \langle \psi ^- |{\mathsf s}^-\rangle
     b_j({\mathsf s})\langle ^+{\mathsf s}|\phi ^+ \rangle
     +\sum_i \int_{-\infty _{II}}^{+\infty}d{\mathsf s}\,
     \langle \psi ^- |{\mathsf s}^-\rangle 
     \langle ^+{\mathsf s}|\phi ^+ \rangle
     \frac{R^{(i)}}{{\mathsf s}-{\mathsf s}_{R_i}} \, . 
     \label{4.6}
\end{equation}
This equality holds for the whole space of functions 
$\langle \psi ^- |{\mathsf s}^-\rangle \langle ^+{\mathsf s}|\phi ^+ \rangle 
\in {\cal S}\cap {\cal H}_-^2$.  Therefore we can omit these arbitrary energy
wave functions 
\begin{equation}
      \langle \psi ^- |{\mathsf s}^-\rangle \langle ^+{\mathsf s}|
     \phi ^+ \rangle =
     \langle \psi ^{out}|{\mathsf s}\rangle \langle {\mathsf s}|
     \phi ^{in}\rangle
     \in {\cal S}\cap {\cal H}_-^2 \, ,
     \label{4.61/2}
\end{equation}
which describe the resolution of the preparation apparatus and of the 
registration apparatus and write equation (\ref{4.6}) as an equation 
between distributions over the function space ${\cal S}\cap {\cal H}_-^2$:
\begin{equation}
\theta({\mathsf s}-m_0^{2})S_j({\mathsf s})=
\theta({\mathsf s}-m_0^{2})b_j({\mathsf s})+
     \sum_i \frac{R^{(i)}}{{\mathsf s}-{\mathsf s}_{R_i}} \, .
      \label{4.7}
\end{equation}
Though one likes to represent the ``physics'' in this apparatus independent
way, what one measures in each experiment contains of course always the 
convolution with an apparatus resolution so that (\ref{4.7})  really means
(\ref{4.6}) in its applications to a particular experiment.  A 
corresponding equation can be written for the partial wave amplitudes 
(\ref{31}) (by dividing (\ref{4.7}) by $2i$ and subtracting $1$ on both
sides for $n=n'$)
\begin{equation}
\theta({\mathsf s}-m_0^{2})a_j({\mathsf s})=
\theta({\mathsf s}-m_0^{2})B_j({\mathsf s})+\sum_i 
      \frac{\tilde{R}^{(i)}}{{\mathsf s}-{\mathsf s}_{R_i}} \, ,
      \label{4.7prime}
\end{equation}
considered as a functional equation in the space of distributions
$\left({\cal S}\cap{\cal H}_{-}^{2}\right)^{\times}$.
Here $B_j({\mathsf s})$ like $b_j(\sm)$ describes an 
ever present slowly varying non-resonant
background.  For instance, if there is only one resonance pole at 
${\mathsf s}_R$ in the $j$-th partial wave, then the cross section contains 
in addition to the resonance part (Breit-Wigner) also interference terms
with the unknown background
\begin{equation}
      |a_j({\mathsf s})|^2=\left| B_j({\mathsf s})+
      \frac{\tilde{R}^{(i)}}{{\mathsf s}-{\mathsf s}_{R_i}} \right| ^2=
      |B_j({\mathsf s})+a_j^{\text{resonance}}({\mathsf s})|^2 \, .
      \label{4.8}
\end{equation}
This shows that it is difficult to distinguish phenomenologically between 
two alternative functions for the resonance part of the amplitude like
(\ref{ajom1}) and (\ref{ajBW2}).  The formula (\ref{4.8}), or the formula 
(\ref{4.7prime}) for a single resonance, is one of the frequently used 
phenomenological formulas.  It also has a theoretical justification under 
the usual analyticity 
assumption for the $S$-matrix $S_j({\mathsf s})$ (Laurent expansion).  

For two (or more) resonances we have obtained (\ref{4.7}), which contains in
addition the interference between two (or more) resonances.  The formula 
(\ref{4.7})
cannot be derived using the usual analyticity assumption of $S_j({\mathsf s})$
only.  We needed for its derivation the Hardy class hypothesis (\ref{2-21}) 
to obtain (\ref{41}) for each pole term {\it separately}.  

The interference of resonances as predicted by (\ref{4.7}), (\ref{4.7prime}) 
has been 
established experimentally for the non-relativistic case in nuclear 
physics and for the relativistic case in the $\rho$-$\omega$ 
interference~\cite{45}.  Since its derivation required the use of the 
assumption 
$\psi ^- \in {\mathbf \Phi} _+$ and $\phi ^+ \in {\mathbf \Phi} _-$ of
Section~\ref{sec:LShardycal}, 
the phenomenological success of formulas
like (\ref{4.7}) for two interfering resonances is another argument in
favour of our new hypothesis of Section~\ref{sec:LShardycal}.

With (\ref{4.7}) and (\ref{4.5a}) we have now established a term-by-term
correspondence between the complex eigenvalue resolution (\ref{4.5a}) of the
prepared in-state vector $\phi ^+$ and the representation 
(\ref{4.7prime}) for the partial wave amplitude (or also (\ref{4.7}) for
the $S$-matrix).  To each Breit-Wigner in (\ref{4.7prime}) corresponds 
a Gamow vector in (\ref{4.2}) or (\ref{4.5a}) and to the background
amplitude $B_j({\mathsf s})$ in (\ref{4.7prime}) corresponds the vector
$\phi ^{bg}$.  This establishes a unique correspondence between the vector
description of quasistable particles and the 
$S$-matrix description of resonances.  
The vector description is used for instance in the 
effective theories with complex Hamiltonian matrix (like the 
Lee-Oehme-Yang theory of $K_S^0$ and $K_L^0$~\cite{LEE} or the finite
dimensional models of nuclear physics~\cite{45}) only that these
finite dimensional models omit the background
vectors $\phi ^{bg}$ (\ref{4.5}) which span an infinite dimensional 
space.  This is the typical feature of the Weisskopf-Wigner approximations. 
In the $S$-matrix description of the resonance by
the phenomenological ansatz (\ref{4.7prime}) one usually does not
omit the background amplitude $B({\mathsf s})$, but often includes in  
$B(\sm)$ also the
contribution of a second distant resonance which according to our 
prediction (\ref{4.7prime}) belongs into the sum of the Breit-Wigner
amplitude.  The
correspondence between (\ref{4.7prime}) and (\ref{4.5a}) unifies the 
theory of resonance scattering and the theory of particle decay.

\section{Summary}
\label{sec:summary}

The mathematical axioms of standard quantum mechanics 
in the Hilbert space are not completely in agreement 
with the calculational tools (like for instance Dirac 
kets) that physicists use in their practical calculations. 
As a consequence of von Neumann's mathematical
assumptions, standard quantum mechanics is time symmetric, 
because the Hilbert Space 
does not allow time asymmetric boundary conditions for the time
symmetric dynamical equations, and the time evolution 
and all other symmetry transformations of non-relativistic
and relativistic space time are described by unitary group
transformations.  
In practice, however, one uses properties of time asymmetry, as is
e.g.\ expressed by the $\pm i\epsilon$ of the Dirac kets 
in the  Lippmann-Schwinger equation.  
Though orthodox quantum theory would require 
that the set of all in-states $\{ \phi^+ \}$ and the set of all 
out-states $\{ \psi^- \}$ is the same Hilbert
space ${\cal H}$, the Lippmann-Schwinger equation suggests a different 
hypothesis.  
If one modifies the Hilbert Space axiom slightly and
replaces it by the new hypothesis that the set of in-states
defined by the preparation apparatus (accelerator) and the set of
out-states or observables defined by the registration apparatus
(detector) form different dense subspaces of the Hilbert Space, 
${\mathbf \Phi}_-$ and ${\mathbf \Phi}_+$, respectively, 
one obtains asymmetric boundary conditions.  
Guided by the Lippmann-Schwinger equations we require that
the space of states ${\mathbf \Phi}_-=\{ \phi^+ \}$ is a Hardy class space 
of the lower complex energy half-plane, and the space 
of observables ${\mathbf \Phi}_+=\{ \psi^- \}$ is a Hardy space of 
the upper complex energy half-plane.  
This assumption allows a mathematically 
rigorous definition of the Dirac-Lippmann-Schwinger kets 
as continuous antilinear functionals,
$|p,\alpha ^+ \rangle \in {\mathbf \Phi}_-^\times$ and
$|p,\alpha ^- \rangle \in {\mathbf \Phi}_+^\times$.
The Hardy class boundary condition 
thus makes the singular Lippmann-Schwinger equation mathematically rigorous.  
But it does more, because in addition to
the Dirac-Lippmann-Schwinger kets for real $\mathsf s$, 
$|{\mathsf s}^- \rangle \in {\mathbf \Phi} _+^{\times}$, the space 
${\mathbf \Phi} _+$
contains also Gamow kets $|{\mathsf s}_R ^-\rangle$, which are generalized 
eigenvectors of the invariant mass operator $P_\mu P^\mu$ with complex
eigenvalue ${\mathsf s}_R=(M_R-i \Gamma _R/2)^2$.  
In addition to the unitary group representations $[j,m^2]$ 
of the Poincar\'e transformation for the time symmetric stationary states 
one obtains also semigroup
representations $[j,{\mathsf s}_R]$ of Poincar\'e transformations 
which distinguish a preferred direction of time.  
Thus the new hypothesis (\ref{2-21}), leads to a time asymmetric 
relativistic quantum theory based, like the time symmetric theory 
for stable particle, on representations of relativistic space-time
transformations.
 
The Gamow kets, spanning the irreducible
representation space $[j, {\mathsf s}_R]$, can be obtained from the
pole position ${\mathsf s}_R=(M_R-i\Gamma _R/2)^2$ of the $j$-th partial 
$S$-matrix $S_j(\mathsf s)$ (or scattering amplitude $a_j(\mathsf s)$) and 
correspond 
to the Breit-Wigner part $a_j^{BW}({\mathsf s})$ of the scattering amplitude.

From the Poincar\'e semigroup transformation of the
Gamow kets, or more specifically from the time evolution 
of the Gamow kets in the rest frame, one obtains the exponential decay law
$e^{-\Gamma _Rt}$ for the Gamow states
$|{\mathsf s}_R^-\rangle$.  
This means that the parameter of exponential time
evolution, the inverse lifetime, is predicted to be
$\hbar/\tau =\Gamma _R=-2{\rm Im}\sqrt{{\mathsf s}_R}$.  
The Gamow vector
unites the notion of resonance described by the Breit-Wigner amplitude
and the notion of decaying particle described by the 
exponential time evolution into
one and the same entity with $\Gamma _R=1/\tau$.  From this we obtain  
the definition of ``mass'' and ``width'' for a relativistic unstable
particle as $M_R={\rm Re}\sqrt{{\mathsf s}_R}$ and 
$\Gamma _R=-2{\rm Im}(\sqrt{{\mathsf s}_R})$ 
and not any of the other definitions of Section~\ref{sec:introduction}, which 
are used for the $Z$-boson in the particle data table \cite{CASO}.

In summary, a relativistic time asymmetric quantum theory that is consistent 
with the Lippmann-Schwinger equation 
leads to a definition of resonance mass $M_R$ and resonance width $\Gamma _R$
for which $1/\Gamma _R$ is the lifetime and for which the $Z$-boson
mass is given by the value $M_R=(91.1626\pm 0.0031) \ {\rm GeV}$, which is
different from either of the two values quoted in the Particle Data 
Table~\cite{CASO} for the $Z$-boson mass.

\acknowledgments

This work was supported by the Welch Foundation and the Humboldt 
Foundation.  Part of this work was done while one of the authors was 
visiting the Institute for Quantum Physics Ulm University and we are 
grateful to Professor W.~Schleich for the hospitality extended to him.  We
also acknowledge valuable discussions on the subject of this paper 
with W.~Blum, W.~Drechsler, N.~Harshman, G.~H\"ohler, T.~Riemann, H.~Saller
and S.~Wickramasekara.

\appendix
\section*{Rigged Hilbert Spaces}
\label{sec:Appendix}

Let ${\mathbf\Phi}_{\text{alg}}$ be a linear space with scalar
product $(\cdot,\cdot)$
(this is the space which most physicists call Hilbert space, but which
mathematicians call a pre-Hilbert space).  Then the triplet of spaces
that form a Rigged Hilbert Space~\cite{38} are obtained by
the completion of ${\mathbf\Phi}_{\text{alg}}$ with respect to three different
topologies (meanings of convergence):
\[
\phi\in{\mathbf\Phi}\subset\HS\subset\Phx\ni\kt{F}.
\]
The space $\HS$ is the mathematician's Hilbert space and it is
complete with respect to the norm convergence $\tau_{\HS}$.  The space
${\mathbf\Phi}$ is complete with respect to a stronger convergence 
$\tau_{\mathbf\Phi}$
(test function space).  The space $\Phx$ is the space of
$\tau_{\mathbf\Phi}$-continuous, antilinear functionals on the space 
${\mathbf\Phi}$:
$\kt{F}\in\Phx$ maps $\phi\in{\mathbf\Phi}$ into ${\mathbb C}$, i.e.\
$F(\phi)=\bk{\phi}{F}\in{\mathbb C}$.  The space $\HS^\times$ of
$\tau_{\HS}$-continuous functionals $f:h\in\HS\rightarrow
f(h)\in{\mathbb C}$ can be identified with $\HS$ by $f(h)=(h,f)$ and
then $\bk{\cdot}{\cdot}$ is the extension of $(\cdot,\cdot)$ to those $F$
which are not elements of $\HS$.

For a $\tau_{\mathbf\Phi}$-\emph{continuous} operator 
$A$ there is also a triplet:
\[
A^\dag|_{\mathbf\Phi} \subset A^\dag \subset A^\times.
\]
Here $A^\dag$ is the Hilbert space adjoint operator and $A^\times$ is
the conjugate operator. $A^\times$ is defined by:
\[
\bk{A\phi}{F}=\br{\phi}A^\times\kt{F},\ \forall\phi
\in{\mathbf\Phi}\ \text{and}\
\forall F\in\Phx.
\]
A vector $\kt{F}\in{\mathbf\Phi}^\times$ is a generalized eigenvector of the
$\tau_{\mathbf\Phi}$-continuous operator $A$ if for some complex number
$\omega$
\[
\bk{A\phi}{F}=\br{\phi}A^\times\kt{F}=\omega\bk{\phi}{F},\
\forall\phi\in{\mathbf\Phi}.
\]
This is also written as $A^\times\kt{F}=\omega\kt{F}$, or
$A\kt{F}=\omega\kt{F}$ if $A^\dag$ is self-adjoint.
An example is the Dirac ket
\[
H^\times\kt{E}=E\kt{E},\ \ E\geq 0.
\]

In the Rigged Hilbert Space one can prove the Nuclear Spectral Theorem
(referring to the nuclearity of the topological spaces ${\mathbf\Phi}$)
which is the most important property for quantum theory and another 
Golden Rule of Dirac: For every self-adjoint operator $H^\dagger$ 
there exists a complete set of generalized eigenvectors 
$|E\rangle\in{\mathbf\Phi}^\times$
such that every vector $\phi\in{\mathbf\Phi}$ can be written as
$$
\phi=\int \rho(E)dE\, |E\rangle\langle E|\phi\rangle \nonumber
$$
where the (wave) functions $\langle E|\phi\rangle\equiv\phi(E)$
are elements of a space of well-behaved functions (e.g.\ the 
Schwartz space of infinitely differentiable rapidly decreasing 
functions). $\rho(E)$ is an integrable function and determines
the normalisation of the eigenvectors
$$
\langle E^\prime|E\rangle=\rho^{-1}(E)\,\delta(E^\prime-E).
$$
The integration can -- in all known cases important for physics --
be chosen to be Riemann integration, and the set of wave functions 
$\phi(E)$ define the space of vectors $\phi\in{\mathbf\Phi}$
and vice-versa.
This theorem generalizes the basis vector expansion in 3- or 
$N$-dimensional spaces
$$
{\vec x}=\sum^{N}_{i=1}{\vec e}_i({\vec e}_i\cdot{\vec x})\quad,
\quad {\vec e}_i\cdot{\vec x}=x^i
$$
to continuously infinite dimensions and is therefore also called
Dirac's basis vector expansion. 
Note that it represents $\phi\in{\mathbf\Phi}$ as a continuous linear 
superposition of basis vectors $|E\rangle\in{\mathbf\Phi}^\times$
which are not elements of ${\mathbf\Phi}$. 
This may appear counter intuitive but has its justification 
in the choice of a set of very special functions $\phi(E)$ as 
possible ``coordinates'' for a vector $\phi\in{\mathbf\Phi}$.

Depending on the choice of the space ${\mathbf\Phi}$, one can also have
generalized eigenvectors with complex eigenvalues for an operator 
$H$ that is essentially
self-adjoint (and bounded from below, which we always assume).
The Gamow vectors are an example:
\[
H^\times\mkt{\zR}=(\zR)\mkt{\zR}.
\]

\end{document}